\documentclass{aa}  
\usepackage{hyperref}
\usepackage{amsmath,amssymb}
\usepackage{txfonts}
\usepackage{graphicx}
\usepackage{bm}
\usepackage{bbold}

\newcommand{\bfq}{\ensuremath{\bm{q}}}
\newcommand{\bfx}{\ensuremath{\bm{x}}}
\newcommand{\bfy}{\ensuremath{\bm{y}}}
\newcommand{\bfz}{\ensuremath{\bm{z}}}

\newcommand{\bbE}{\ensuremath{\mathbb{E}}}
\newcommand{\bbR}{\ensuremath{\mathbb{R}}}

\DeclareSymbolFont{bbold}{U}{bbold}{m}{n}
\DeclareSymbolFontAlphabet{\mathbbold}{bbold}
\newcommand{\bbone}{\ensuremath{\mathbbold{1}}}

\newcommand{\CB}{\ensuremath{B}}
\newcommand{\CW}{\ensuremath{W}}

\newcommand{\DD}{\ensuremath{\textsf{DD}}}
\newcommand{\DR}{\ensuremath{\textsf{DR}}}
\newcommand{\RR}{\ensuremath{\textsf{RR}}}

\newcommand{\DRt}{\ensuremath{\mathcal{DR}}}
\newcommand{\RRt}{\ensuremath{\mathcal{RR}}}

\newcommand{\Vt}{\ensuremath{\mathcal{V}}}
\newcommand{\Vest}{\ensuremath{\textsf{V}}}
\newcommand{\Ut}{\ensuremath{\mathcal{U}}}
\newcommand{\Uest}{\ensuremath{\textsf{U}}}

\newcommand{\Nr}{{\ensuremath{N_\mathrm{r}}}}
\newcommand{\Nq}{{\ensuremath{N_\mathrm{q}}}}
\newcommand{\Nsh}{{\ensuremath{N_{\text{sh}}}}}

\newcommand{\rmd}{\ensuremath{\mathrm{d}}}
\newcommand{\at}{\ensuremath{\tan^{-1}}}
\newcommand{\mDelta}{\ensuremath{\delta}}
\newcommand{\hMpc}{\ensuremath{\text{Mpc}/h}}

\begin{document}

\title{Improving the accuracy of estimators for the two-point correlation
  function }
\author{Martin Kerscher}
\institute{Ludwig--Maximilians Universt\"at M\"unchen, 
  Fakult\"at f\"ur Physik, Schellingstr. 4, D-80799 M\"unchen\\ 
  \email{martin.kerscher@lmu.de} }
\date{August 25, 2022}
 
\abstract
    {}
    {We show how to increase the accuracy of estimates of the two-point
      correlation function without sacrificing efficiency.}
    {We quantify the error of the pair-counts and of the
      Landy\,\&\,Szalay estimator by comparing them with exact
      reference values.  The standard method, using random point sets,
      is compared to geometrically motivated estimators and estimators
      using quasi-Monte~Carlo integration.}
    {In the standard method, the error scales proportionally to
      $1/\sqrt{\Nr}$, with $\Nr$ being the number of random points.
      In our improved methods, the error scales almost proportionally
      to $1/\Nq$, where $\Nq$ is the number of points from a
      low-discrepancy sequence.
      We study the runtimes of the new estimator in comparison to those of the
      standard estimator, keeping the same level of accuracy.  For the
      considered case, we always see a speedup ranging from 50\% up to a
      factor of several thousand.
      We also discuss how to apply these improved estimators
      to incompletely sampled galaxy catalogues.}
   {}

   \keywords{Methods: statistical, data analysis, Cosmology:
     large-scale structure of Universe}

\maketitle

\section{Introduction}  

Statistical approaches are often used to characterise the large-scale
structure of the galaxy distribution, wherein it is assumed that the
distribution of galaxies is a realisation of a point process (see
e.g.\,\citealt{neyman:statistical};
\citealt{peebles:lss}). Observations give us the positions of galaxies
in space.  From this set of points, we estimate the moments of the
point process, specifically the two-point correlation function
$\xi(r)$ or the two-point density
\begin{align}
  \label{eq:defxi}
  \varrho_2(\bfx,\bfy) =  \varrho^2\ \left(1 + \xi(|\bfx-\bfy|)\right),
\end{align}
the probability of finding two galaxies at $\bfx$ and
$\bfy$, where $\varrho$ is the number density. In a homogeneous and
isotropic point process, $\xi(r)$ only depends on the separation
$r=|\bfx-\bfy|$.
To determine $\xi(r)$ from a galaxy catalogue within a finite domain
$\CW\subset\bbR^3$, we use estimators.  In cosmology, estimators based
on random point sets are most commonly used. These rely on the
data--data $\DD$, data--random $\DR$, and random--random $\RR$
pair-counts (see below for the definition).

The two-point correlation function of the galaxy distribution is often
used to constrain models of structure and galaxy formation and to
estimate parameters of cosmological models.  In current and upcoming
galaxy samples, the positions of millions and up to billions of galaxies
will be observed
(\citealt[BOSS]{dawson:boss}, \citealt[eBOSS]{ross:complete}, 
\citealt[DES]{abbott:des3}, \citealt[DESI]{aghamousa:desi}, 
\citealt[PAUS]{alarcon:pau}, \citealt[LSST]{ivezic:lsst}, 
\citealt[Euclid]{amendola:euclid}).
Fast and reliable methods for calculating the two-point correlation
function are needed. These large samples allow us to reduce the
statistical error from cosmic variance, but for the error budget of
the two-point correlation function we also have to control the
systematic errors. One systematic contribution is the error from
random sets used in the pair-counts $\DR$ and $\RR$.
As an example, consider the baryon acoustic oscillations (BAOs) which
lead to a peak in the two-point correlation function of galaxies at a
scale of about 100\hMpc\ (\citealt{eisenstein:baryon};
\citealt{bautista:complete}). This BAO peak has a height of
approximately 0.01 above zero (compare Fig.\,2 in
\citealt{eisenstein:baryon}). For a percent level accuracy, we need to
calculate the two-point correlation function with an absolute accuracy
of less than $10^{-4}$.
Below, we show how to reduce the systematic error to this level of
accuracy without sacrificing efficiency.

Several estimators for the two-point correlation function have been
developed (\citealt{peebles:statisticalIII};
\citealt{hewett:estimation}; \citealt{davis:surveyV};
\citealt{rivolo:two-point}; \citealt{landy:bias};
\citealt{hamilton:towards}). By comparing these estimators to a
reference result from a cosmological simulation,
\citet{kerscher:comparison} found that the \citet{landy:bias}
estimator is the preferred estimator with the smallest deviation from
the reference and also that its bias is negligible compared to its
variance.

We focus on methods for increasing the numerical accuracy of the
pair-counts as used in these estimators.
The random point set, shared in the pair-counts $\RR$ and $\DR$ is
used to correct for boundary (finite-size) and inhomogeneous sampling
effects. As we outline below, $\RR$ and $\DR$ are Monte~Carlo volume
integration schemes. As expected from standard Monte~Carlo
integration, the error of these pair-counts $\RR$ and $\DR$ scales at
least as $1/\sqrt{\Nr}$, where $\Nr$ is the number of random points
used (see Sect.\,\ref{sec:RR} for a more differentiated view). This
slow convergence rate makes increasing the accuracy costly, and sometimes
unfeasible.
To improve the accuracy of the pair-counts without sacrificing
efficiency, we follow two directions:
\begin{itemize}
\item The pair-counts can be expressed as averages of specific volume
  fractions. We use this to propose special adapted volume integration
  schemes which, in turn, can be calculated more efficiently than the
  standard approach.
\item We replace the standard Monte~Carlo scheme with a
  quasi-Monte~Carlo integration, which leads to an improved scaling of
  the error that is almost proportional to $1/\Nq$, where $\Nq$ is the
  number of points from a low-discrepancy sequence.
\end{itemize}
We compare the standard and the new methods to exactly known reference
values. This allows us to empirically validate the asymptotic scaling
of the errors.

A variety of approaches have been suggested to improve the
speed and accuracy of estimators for the two-point correlation
function.
\citet{keihaennen:estimating} show that at fixed computational cost, a
split random catalogue improves the accuracy of estimators for the
two-point correlation function.  For galaxy catalogues,
\citet{demina:computationally} achieve a speedup by factorising the
calculations in radial (redshift) and angular coordinates (see also
\citealt{breton:fast}).
Perhaps closest to our work are the investigations by
\citet{davila-kurban:improved}. These authors use glass-like point
sets instead of the random catalogues, where we use low-discrepancy
sequences.
As reference values in our comparisons, we use the exact results from
\citet{baddeley:3dpoint} and \citet{kerscher:twopoint} for a
rectangular box as summarised in Appendix\,\ref{sec:simple-win} (for
periodic boxes see Appendix\,\ref{sec:periodic}).  \citet{he:fast}
also discusses some approximations for these exact results.

Other approaches focus on the computational problem of calculating the
pair-counts.  Tree-based methods can be significantly faster than a
direct implementation of the pair-counts, specifically for small radii
(\citealt{moore:fast}, see also Appendix\,\ref{sec:implementation}).
The double loop in the pair-count calculations can be parallelised.
\citet{alonso:cute} showed how to obtain a speedup by a factor of 100
over the direct implementation by using multi-threading on multi-core
CPUs or utilising many cores in GPUs.
It is well known from matrix computations that the memory layout of
the data can have dramatic consequences for the runtime of algorithms
(see e.g.\,\citealt{anderson:lapack}).  Also, for the pair-counts, a
clever layout of the coordinates in the memory can lead to a
significant speedup \citep{donoso:gundam}. A similar approach can be
combined with multi-threading and vectorisation resulting in a
blazingly fast code \citep{sinha:corrfunc}.
Our conceptual improvements can be combined with these 
computational speedups.

In Sect.\,\ref{sec:stuffweneed} we give the definition of the
pair-counts and discuss the geometry of the expected pair-counts. In
Appendix\,\ref{sec:expectpair} we give the details of the derivations
and in Appendix\,\ref{sec:simple-win} we summarize some results for
simple sample geometries.  Together with Sect.\,\ref{sec:stuffweneed},
this enables us to calculate exact reference values for the
pair-counts.
At the end of Sect.\,\ref{sec:stuffweneed}, we give a short introduction
to the quasi-Monte~Carlo method, as used in our improved estimators.
In Sects.\,\ref{sec:RR} and \ref{sec:DR}, we compare the standard and
improved versions of the pair-counts with the exactly known reference
values. We discuss the scaling of the error with the number of points.
We put this together in Sect.\,\ref{sec:xi} and show how an improved
version of the \citet{landy:bias} estimator can be constructed. Again
we discuss the scaling of the error with the number of points.
In Sects.\,\ref{sec:runtime} and \ref{sec:dense}, we compare the run
times of the new estimator with those of the standard
\citet{landy:bias} estimator in some typical situations.
In Sect.\,\ref{sec:incomplete}, we discuss how these improvements have
to be adapted to estimate the two-point correlation function from an
inhomogeneous sampled galaxy distribution.
We summarise in Sect.\,\ref{sec:summary} and give some recommendations.
In Appendix\,\ref{sec:implementation}, we discuss details of the 
implementation, the run times, and give a link to the code.

\section{Pair-counts, geometry, and
  quasi-Monte~Carlo}
\label{sec:stuffweneed}

The set of the $N$ data points (e.g.~galaxies) is $\{\bfx_i\}_{i=1}^N$
with all points $\bfx_i\in\CW$ inside the observation window
$\CW\subset\bbR^3$, that is,\ inside the unmasked area.  The number
density is estimated with $\widehat{\rho}=\tfrac{N}{|\CW|}$, where
$|\CW|$ is the volume of $\CW$.  We then define
\begin{equation}
\label{eq:defDD}
\DD(r) = \frac{1}{N^2} \sum_{i=1}^N\sum_{j=1, j\ne i}^N\
\mDelta k_r^\mDelta(|\bfx_i-\bfx_j|),
\end{equation}
the normalised number of data--data pairs with a distance of
$r=|\bfx_i-\bfx_j|$ in the interval $[r,r+\mDelta]$. We use a
rectangular kernel
\begin{equation}
  k_r^\mDelta(s) = \tfrac{1}{\mDelta}\mathbb{1}_{[r,r+\mDelta]}(s),
\end{equation}
with the indicator function of the set $A$ defined as
\begin{equation}
\mathbb{1}_A(q) =
\begin{cases}
  1 & \text{if } q\in A,\\
  0 & \text{else}.
\end{cases}
\end{equation}
Also, other kernels with $\int k_r^\mDelta(s)\rmd s=1$ are possible
(e.g.\ triangular, truncated Gaussian, or Epanechnikov).
We consider $\Nr$ randomly distributed points
$\{\bfy_j\}_{j=1}^{\Nr}$, all inside the sample geometry
$\bfy_j\in\CW$.  The normalised number of data-random pairs with a
distance in $[r,r+\mDelta]$ is denoted by
\begin{equation}
  \label{eq:defDR}
  \DR(r) =  \frac{1}{N \Nr}  \sum_{i=1}^N\ \sum_{j=1}^{\Nr}\
  \mDelta k_r^\mDelta(|\bfx_i-\bfy_j|).
\end{equation}
Similarly,
\begin{equation}
  \label{eq:defRR}
  \RR(r) = \frac{1}{\Nr^2}\sum_{i=1}^{\Nr} \sum_{j=1, j\ne i}^{\Nr}\
  \mDelta k_r^\mDelta(|\bfy_i-\bfy_j|),
\end{equation}
is the normalised number of random--random pairs.
The \citet{landy:bias} estimator is defined as
\begin{align}
  \label{eq:xiLS}
\xi_\text{LS}(r) &= \frac{\DD(r) - 2\DR(r) + \RR(r)}  {\RR(r)} .
\end{align}
Also, the estimators provided by \citet{peebles:statisticalIII},
\citet{hewett:estimation}, \citet{davis:surveyV}, and
\citet{hamilton:towards} can be defined in terms of the pair-counts
and our results apply accordingly.

\subsection{Geometry of pair-counts}
\label{sect:geometrypaircount}

The expectation of the pair-counts $\DR$ and $\RR$ can be expressed in
terms of geometric quantities depending on the sample window $\CW$ and
on the point set (for $\DR$, \citealt{kerscher:twopoint}).
We first consider the set-covariance
\begin{align}
  \label{eq:setcov}
  \gamma_\CW(\bfx) &= |\CW \cap \CW_{\bfx}|
  = \int_{\bbR^3}\bbone_\CW(\bfy)\bbone_\CW(\bfy+\bfx) \rmd \bfy,
\end{align}
where $\CW_{\bfx}$ is the shifted window $\CW$, that is,\ the set of
all points from $\CW$ shifted by the vector $\bfx$.
$|\CW \cap \CW_{\bfx}|$ is the volume of the set $\CW\cap\CW_{\bfx}$.
The isotropised set-covariance $\overline{\gamma_\CW}(r)$ can be
calculated from $\gamma_\CW(\bfx)$:
\begin{equation}
\label{eq:isotrop-setcov}
\overline{\gamma_\CW}(r) = \frac{1}{4\pi}\int_0^\pi\int_0^{2\pi}\,
\gamma_\CW(\bfx(r,\theta,\phi))\ \sin(\theta)\rmd\theta\rmd\phi.
\end{equation}
Here $\bfx(r,\theta,\phi)=(r\cos(\phi)\sin(\theta),
r\sin(\phi)\sin(\theta), r\cos(\theta))$.
For a large number of random points $\Nr$  one
obtains (see Appendix\,\ref{sec:expectpair}):
\begin{align}
  \label{eq:expectRR}
  \RR(r) \rightarrow\ \RRt(r)
  &= \frac{4\pi}{|\CW|^2} \int_{r}^{r+\mDelta}\!\!
  \overline{\gamma_\CW}(s)\ s^2\rmd s\\
  &\approx \frac{4\pi r^2 \mDelta}{|\CW|^2}\ \overline{\gamma_\CW}(r)
  \quad \text{for } \mDelta \text{ small}. \nonumber
\end{align}
where $\RRt(r)$ is the expectation value of the pair-count $\RR(r)$
illustrating its geometric nature. 

\citet{ripley:spatial} used a local area weight in an estimator for
his $K$-function (the normalised integrated two-point density) and
\citet{rivolo:two-point} considered a similar weight in his estimator
for the two-point correlation function.
This weight is inversely proportional to the fraction of the surface
area of a sphere $\CB_{r}(\bfy)$ with radius $r$ centred on the point
$\bfy$ inside $\CW$:
\begin{equation}
  \text{area}(\partial\CB_{r}(\bfy)\cap\CW)
  = \int_0^\pi\int_0^{2\pi}\, \bbone_\CW(\bfy+\bfx(r,\theta,\phi))\
  \sin(\theta)\rmd\theta\rmd\phi.
\end{equation}
For a large number of random points $\Nr$, one obtains (see
Appendix\,\ref{sec:expectpair}):
\begin{align}
  \label{eq:expectDR}
  \DR(r) \rightarrow\ \DRt(r)
  & = \frac{1}{|\CW|N}\sum_{i=1}^N \int_r^{r+\mDelta}\!\!\!\!\!
    \text{area}(\partial\CB_{s}(\bfx_i)\cap\CW)\,\rmd s\\ 
  & \approx \frac{1}{|\CW|N}
  \sum_{i=1}^N  \text{area}(\partial\CB_{r}(\bfx_i)\cap\CW)\,\mDelta,
  \ \text{for } \mDelta \text{ small}. \nonumber
\end{align}
As before, $\DRt(r)$ is the expectation value of the pair-count
$\DR(r)$ illustrating its geometric nature. However, now both $\DR(r)$
and $\DRt(r)$ are depending on the points $\{\bfx_i\}_{i=1}^N$ under
consideration.

In Appendix\,\ref{sec:simple-win}, we give expressions for
$\overline{\gamma_\CW}(r)$ and
$\text{area}(\partial\CB_{r}(\bfy)\cap\CW)$ if $\CW$ is a rectangular
box or a sphere. From these expressions, we calculate the reference
values $\DRt$ and $\RRt$. This allows us to compare different
integration schemes for the pair-counts $\DR$ and $\RR$ and we can
investigate the scaling of the accuracy with the number of points used
in these methods.

\subsection{Quasi-Monte~Carlo}
\label{sec:qmc}

In our improved methods for estimating the pair-counts, we use
quasi-Monte~Carlo integration.
In a standard Monte~Carlo integration scheme, one uses $\Nr$ random
points $\{\bfy_1,\ldots,\bfy_\Nr\}$ to estimate the integral
$\int_{[0,1]^d} f(\bfx)\mathrm{d}\bfx$ by
$\frac{1}{\Nr} \sum_{i=1}^\Nr f(\bfy_i)$.  The accuracy can be
estimated using the Chebyshev inequality, which tells us that the
probability of an error exceeding a given threshold decreases with
$1/\!\sqrt{\Nr}$. In other words, the standard error of a Monte~Carlo
integration scales as $1/\!\sqrt{\Nr}$

With quasi-Monte~Carlo methods, we use
$\frac{1}{\Nq} \sum_{i=1}^\Nq f(\bfq_i)$ to numerically integrate
$\int_{[0,1]^d} f(\bfx)\mathrm{d}\bfx$ (see e.g.\
\citealt{niederreiter:random}). This estimate of the integral almost
looks identical to the Monte~Carlo integration above. However, for a
quasi-Monte~Carlo integration, the points in
$Q=\{\bfq_1,\ldots,\bfq_\Nq\}$ are not random.  It is essential for
the application of quasi-Monte~Carlo integration that for a given
point set $Q$ the error bound
\begin{equation}
\label{eq:koksama}
\left| \int_{[0,1]^d} f(\bfx)\mathrm{d}\bfx  -
  \frac{1}{\Nq} \sum_{i=1}^\Nq f(\bfq_i)\right| \le V(f)\, D(Q)
\end{equation}
factorises into a measure of variation $V(f)$   depending only on
properties of $f$, and a measure of discrepancy $D(Q)$  
depending only on the properties of the point set $Q$.
If we consider functions of bounded variation,
equation\,(\ref{eq:koksama}) is referred to as the Koksama-Hlawka
bound and the measure of discrepancy is the star discrepancy (see
e.g.\,\citealt{lecuyer:recent}).
To control the error bound\,(\ref{eq:koksama}), we have to control
$D(Q)$ (we note that $V(f)$ does not depend on the point set
$Q$). Low-discrepancy sequences, such as the Halton sequence, have
been constructed with that in mind. For such sequences,
{}\cite{halton:efficiency} showed that
\begin{equation}
D(Q) \propto \frac{(\log \Nq)^d}{\Nq} .
\label{eq:discrepancyscaling}
\end{equation}
For small dimensions, $d,$ this compares favourably to a Monte~Carlo
integration where the standard error only scales as $1/\!\sqrt{\Nr}$.
Halton sequences can be used to estimate integrals over indicator
functions, which in turn define volumes like the set covariance. As
indicator functions have bounded total variation,
Eq.\,(\ref{eq:koksama}) applies.

Upper bounds like Eq.\,(\ref{eq:koksama}) are worst-case bounds.
\cite{owen:strong} derive an analogue to a strong law of large numbers
for randomised low-discrepancy sequences.  This further justifies the
procedure for scrambling the Halton sequence developed by
\cite{owen:randomized}, where the scaling from
Eq.\,(\ref{eq:discrepancyscaling}) in Eq.\,(\ref{eq:koksama}) still
gives an upper bound, but on average smaller errors are expected.  We
use these randomised Halton sequences in our calculations (see
Appendix\,\ref{sec:implementation}).

\section{$\RR$}
\label{sec:RR}

First we investigate how the accuracy of the standard $\RR$, as given
in Eq.\,(\ref{eq:defRR}), scales with the number of random points
used. The numerical implementation of the pair-counts is discussed
in Appendix.\,\ref{sec:implementation}.
The expectation value $\RRt$ of $\RR$ can be expressed in terms
of the isotropised set-covariance; see Eq.\,(\ref{eq:expectRR}). Using
the Eqs.\,(\ref{eq:setcovbox}) and (\ref{eq:int_isosetcovbox}) from
Appendix\,\ref{sec:setcov-afrac-box} for the isotropised set
covariance, we calculate $\RRt(r)$ as a reference value for rectangular
boxes $W$.
\begin{figure}
\begin{center}
\includegraphics[width=0.38\textwidth]{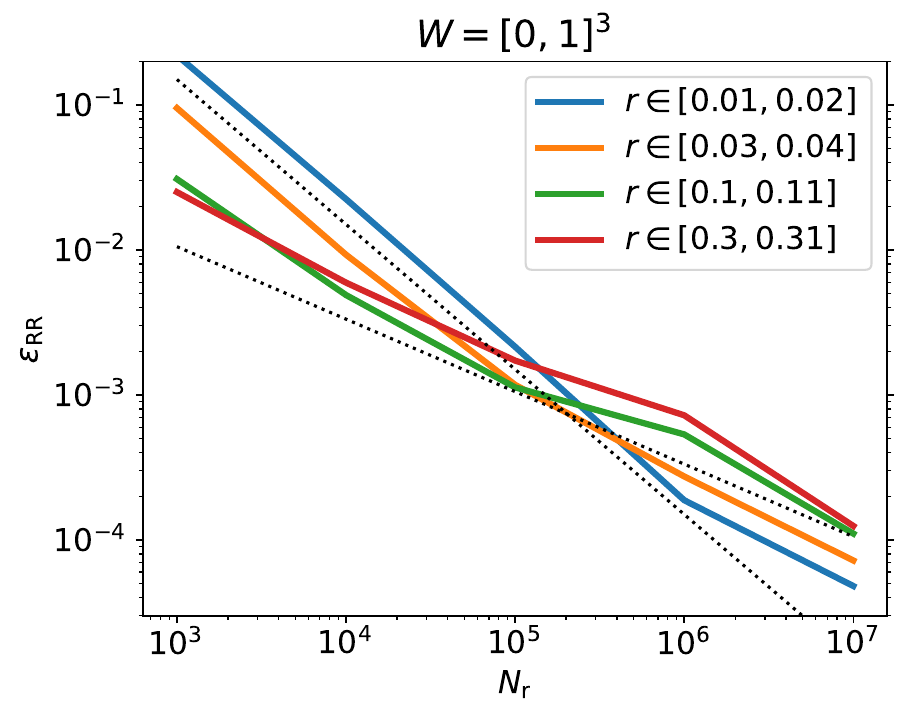}\\
\includegraphics[width=0.38\textwidth]{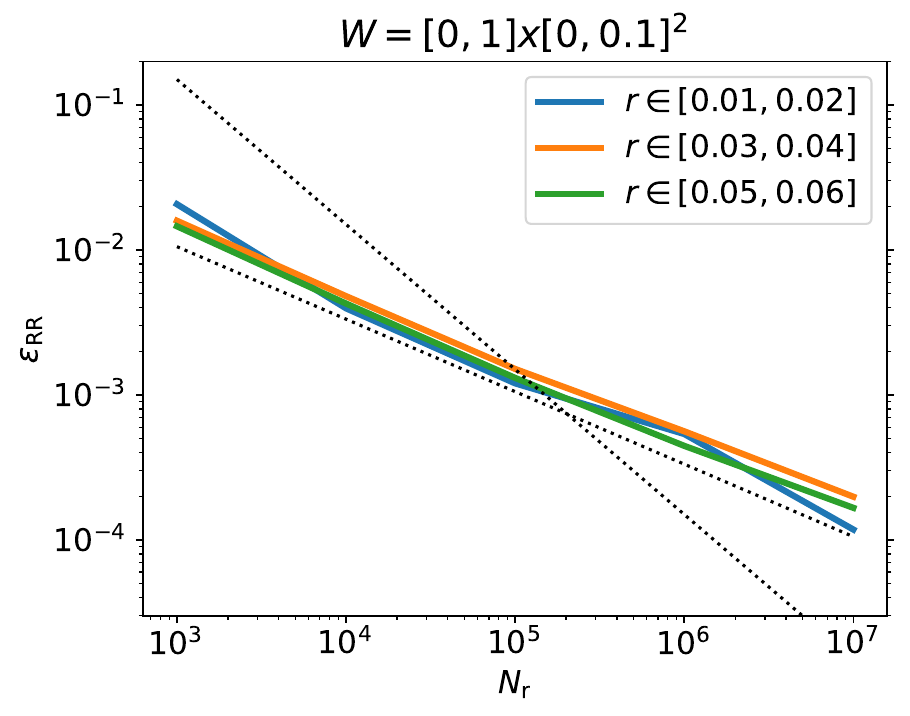}
\end{center}
\caption{Relative error $\varepsilon_{\RR}$ against the number of
  random points $\Nr$ used in the standard procedure for calculating
  $\RR(r)$ in rectangular windows $\CW$.  The black dotted lines are
  proportional to $1/\!\Nr$ and $1/\!\sqrt{\Nr}$.}
\label{fig:RRscaling}
\end{figure}
Figure~\ref{fig:RRscaling}  shows the scaling of the relative error
$\varepsilon_{\RR}=|\RRt(r)-\RR(r)|/\RRt(r)$ with the number of random
points used. 
The error $\varepsilon_{\RR}$ is the mean value calculated from
100\,samples of the random point sets. Only for the largest $\Nr$ do we
use 10 samples. This also applies to the errors calculated in
the following sections.
We show the results for a rectangular cuboid $W=[0,1]^3$ and for a
thin rectangular box $W=[0,1]\times[0,0.1]^2$. With the thin box, we
investigate how the error is affected in a sample $W$ where the
boundary effects are more dominant.
\citet[their eq.\,(21)]{landy:bias} show that the relative error of
$\RR$ is of the form $\frac{A}{\sqrt{\Nr}}+\frac{B}{\Nr}$, with $A$
and $B$ depending on the window~$W$ and the bin width.  As expected,
we see in Fig.\,\ref{fig:RRscaling} that the error scales as
$1/\!\sqrt{\Nr}$ for large $\Nr$. For small $\Nr$ and small radii, the
contribution proportional to $1/\Nr$ appears. This additional
contribution can be seen in the voluminous sample but not in the thin
box.
Simply using a three-dimensional low-discrepancy sequence instead 
of the random points in Eq.\,(\ref{eq:defRR}) is not feasible, as we 
see below.

From Eqs.\,(\ref{eq:S-campbel-mecke}) and (\ref{eq:RRt}) we get
\begin{equation}
  \label{eq:RR6d}
  \RRt(r) = \frac{1}{|\CW|^2} \int_{\mathbb{R}^3} \int_{\mathbb{R}^3}
  \mathbb{1}_\CW(\bfy)\mathbb{1}_\CW(\bfz)\,
  \mDelta k_r^\mDelta(|\bfy-\bfz|) \, \rmd\bfy\,\rmd\bfz.
\end{equation}
This allows a more flexible approach.  Consider two (random) point
sets $P_1=\{\bfy_i\}_{i=1}^{\Nr}$ and $P_2=\{\bfz_j\}_{j=1}^{\Nr}$,
with the points $\bfy_i\in\CW$, $\bfz_j\in\CW$ all inside the sample
geometry. We then define
\begin{align}
  \label{eq:RR2x3d}
  \RR_{2x3d}(r) &= \frac{1}{\Nr^2}\sum_{i=1}^{\Nr}
  \sum_{j=1}^{\Nr}\ \mDelta k_r^\mDelta(|\bfy_i-\bfz_j|),
\end{align}
which is also an estimate of $\RRt(r)$.  
If the points in $P_1$ and $P_2$ are drawn from a Poisson process, the
estimate from Eq.\,(\ref{eq:RR2x3d}) is almost the same as the
estimate from Eq.\,(\ref{eq:defRR}), because points in a Poisson
process are independent. See also \cite{davila-kurban:improved}, who
observe that one cannot use $P_1\equiv P_2$ for $\RR_{2x3d}$ when they
construct an estimator for the two-point correlation function using
`glass like' point sets.
The representation of $\RRt$ in Eq.\,(\ref{eq:RR6d}) as a
six-dimensional integral suggests a six-dimensional (quasi-)Monte~Carlo
approach.  Consequently, we use six-dimensional random points or a
six-dimensional randomised Halton sequence $\{\bfq_i\}_{i=1}^{\Nq}$,
which we split as $\bfq_i=(\bfy_i,\bfz_i)$ into two three-dimensional
sequences. We scale the points in the sequences
$\{\bfy_i\}_{i=1}^{\Nq}$ and $\{\bfz_i\}_{i=1}^{\Nq}$ such that each
$\bfy_i\in\CW$ and $\bfz_j\in\CW$ are uniformly distributed inside the
sample geometry.

\begin{figure}
\begin{center}
\includegraphics[width=0.38\textwidth]{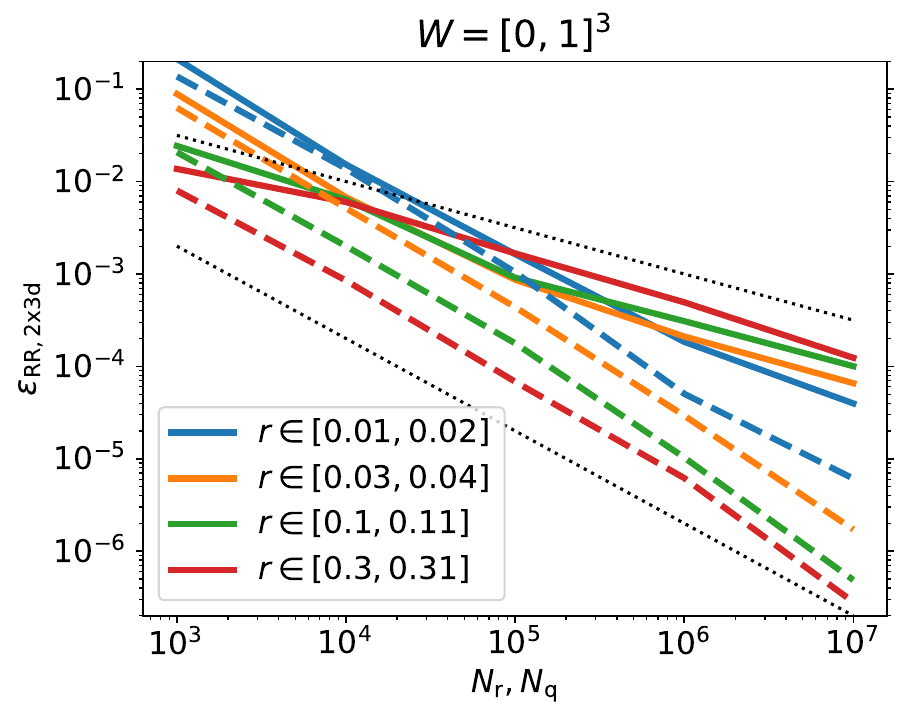}
\includegraphics[width=0.38\textwidth]{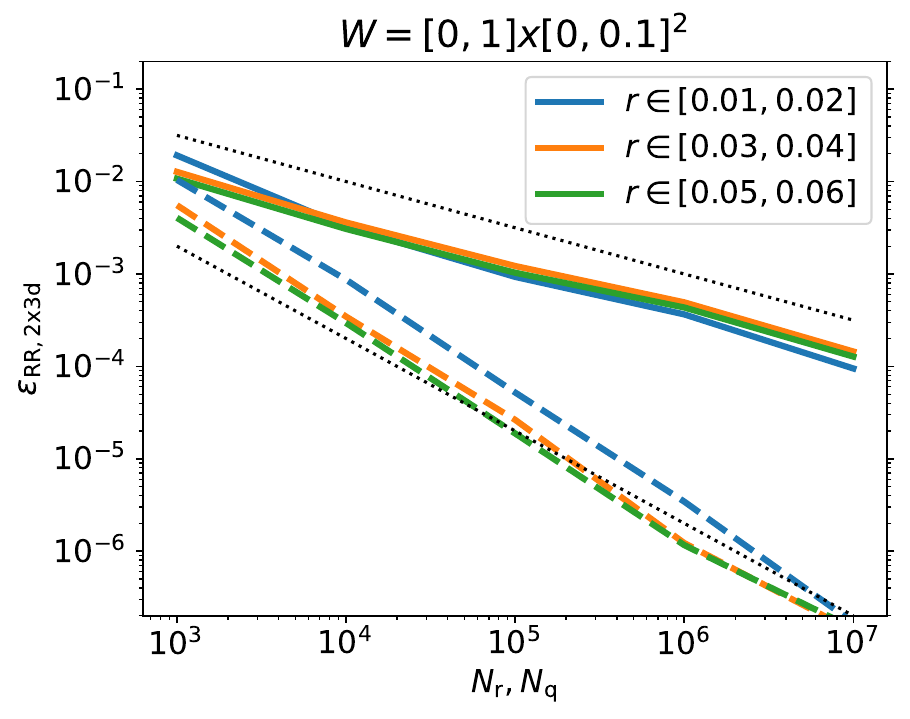}
\end{center}
\caption{Relative error $\varepsilon_{\RR,2x3d}$ calculated for
  rectangular windows $\CW$ with a pure Monte~Carlo integration (solid
  lines) and with a randomised Halton sequence (dashed lines). The
  black dotted lines are proportional to $1/\!\sqrt{\Nr}$ (upper) and
  $1/\Nq$ (lower).  }
\label{fig:RR2x3dscaling}
\end{figure}

Figure\,\ref{fig:RR2x3dscaling}  compares the scaling of the relative
error $\varepsilon_{\RR,2x3d}=|\RRt(r)-\RR_{2x3d}(r)|/\RRt(r)$ with
the number of (quasi-)random points.
For random points, we see the expected scaling $1/\!\sqrt{N}$ of the
the error.  Using the quasi-Monte~Carlo approach, we see that the error
scales proportionally to $1/N$, which is even faster than the
theoretical expectation according to
Eq.\,(\ref{eq:discrepancyscaling}).  Using a low-discrepancy sequence,
we gain more than two orders of magnitude in accuracy compared to
the random point sets.

\section{$\DR$}
\label{sec:DR}

In Sect.\,\ref{sect:geometrypaircount} we show that the expectation of
$\DR(r)$ is $\DRt(r)$ and that it can be calculated from the area
fraction $\text{area}(\partial\CB_{r}(\bfx_i)\cap\CW)$.
For a rectangular box, \cite{baddeley:3dpoint} gave explicit
expressions for $\text{area}(\partial\CB_{r}(\bfx_i)\cap\CW)$ (see
Appendix\,\ref{sec:setcov-afrac-box}). We use these expressions
to calculate $\DRt(r)$ according to Eqs.\,(\ref{eq:Vrdelta}) and
(\ref{eq:DRt-area}).
In the numerical integration of Eq.\,(\ref{eq:Vrdelta}), we make sure
to achieve a relative error of at least $10^{-10}$ for our reference
value $\DRt(r)$.
Both $\DR(r)$ and $\DRt(r)$  depend on the point set under
consideration. We need a realistic data set
$\textsf{D}=\{\bfx_i\}_{i=1}^N$ to calculate both the theoretical
reference $\DRt(r)$ and the different estimates for $\DR(r)$. For this
purpose, we use a sample of simulated galaxy
clusters\footnote{Specifically we use the simulated galaxy clusters
  from the snapshot \texttt{Box2/hr, snap\_136, z=0.066340191}
  downloaded from
  \url{http://www.magneticum.org/data.html\#FULL_CATALOUGES}.}  from
the Magneticum simulation
(\citealt{hirschmann:cosmological}; \citealt{ragagnin:web}).
The side length of the simulation box is $325\,\hMpc$ and we use the
real space positions of the 10429 simulated clusters.  We rescale the
coordinates by the side-length of the box such that all points are
inside $\CW=[0,1]^3$. In the smaller window $\CW=[0,1]\times[0,0.1]^2$
only 86 clusters are left.

\begin{figure}
\begin{center}
\includegraphics[width=0.38\textwidth]{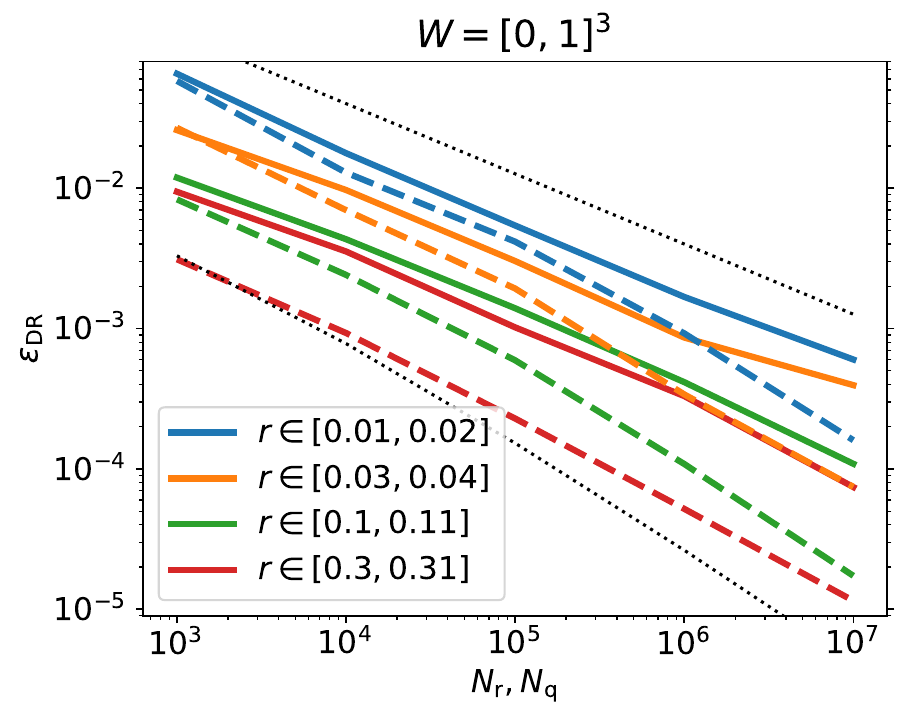}
\includegraphics[width=0.38\textwidth]{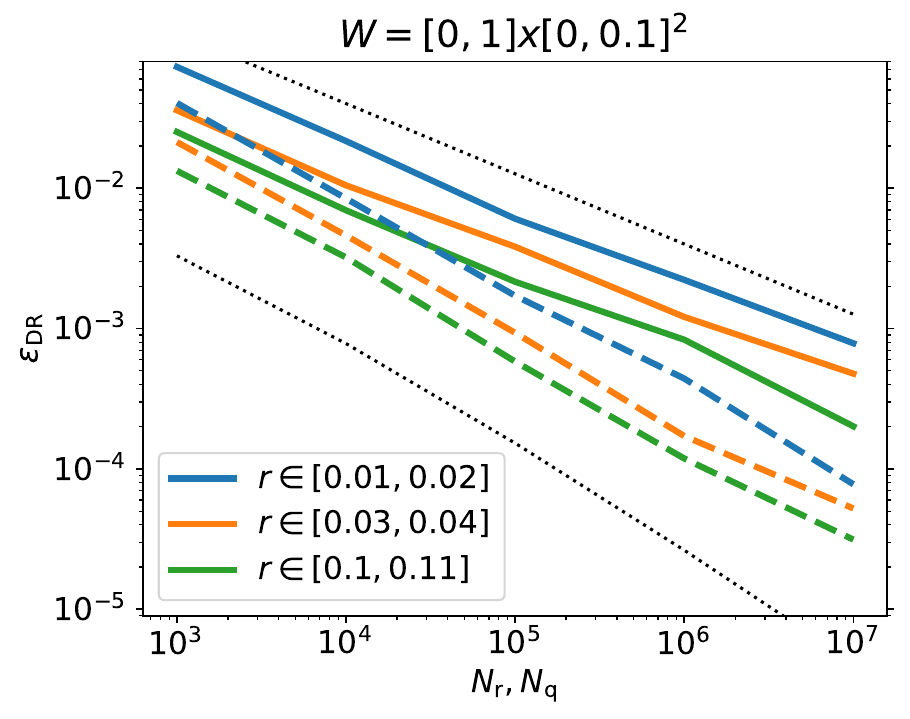}
\end{center}
\caption{
  \label{fig:epsDR}
  Relative error $\varepsilon_{\DR}$ calculated for rectangular
  windows $\CW$ with a standard Monte~Carlo integration (solid lines)
  and with a quasi-Monte~Carlo scheme using a randomised Halton
  sequence (dashed lines). The black dotted lines are proportional to
  $1/\!\sqrt{N}$ (upper) and $(\log N)^3/N$ (lower).}
\end{figure}

Now we are set to determine the scaling of the relative error
$\varepsilon_{\DR}=|\DRt(r)-\DR(r)|/\DRt(r)$ with the number of
(quasi-)random points used. Figure\,\ref{fig:epsDR} compares the
standard approach with ordinary random numbers in $\DR(r)$ to a
pair-count determined using a low-discrepancy sequence instead of the
random points.  Using a randomised Halton sequence, we only observe a
minor gain for small $N$, but for large $N$ the error is reduced by an
order of magnitude. In the small window, $W=[0,1]\times[0,0.1]^2$,
this is more pronounced. The scaling follows the expected behaviour
from Eq.\,(\ref{eq:discrepancyscaling}).

We can do better if we consider the geometry of $\DR(r)$. From
Eq.\,(\ref{eq:Vrdelta}) and Eq.\,(\ref{eq:DRt-area}) we know that
the expectation value of $\DR(r)$ is 
$\DRt(r)=\frac{1}{|\CW|N}\sum_{i=1}^N\Vt_r^\mDelta(\bfx_i)$, where
\begin{align}
  \Vt_r^\mDelta(\bfx_i) 
  & = \int_r^{r+\mDelta}\!\!\!\!\!
    \text{area}(\partial\CB_{s}(\bfx_i)\cap\CW)\,\rmd s
  = |S_r^\mDelta(\bfx_i)\cap\CW|
\end{align}
is the volume of the spherical shell with a radial range in
$[r,r+\mDelta]$ around $\bfx_i$ inside the sample geometry $\CW$.  As
already suggested by \citet{rivolo:two-point}, this directly leads to
a Monte~Carlo scheme.
With $\Nsh$ points $\{\bfy_i\}_{i=1}^{\Nsh}$ (quasi-)randomly
distributed in the
shell 
$S_r^\mDelta(\bfx_i)=\{\bfy\in\bbR^3 \,|\, s < |\bfy-\bfx_i|\le
s+\mDelta \}$ around
$\bfx_i$ we define a (quasi-)Monte~Carlo estimate of
$\Vt_r^\mDelta(\bfx_i):$
\begin{align}
  \label{eq:Vest}
  \Vest_s^\mDelta(\bfx_i) &=  \frac{|S_r^\mDelta|}{\Nsh}
  \sum_{j=1}^{\Nsh}\mathbb{1}_\CW(\bfy_j), 
\end{align}
with the volume
$|S_r^\mDelta|=\frac{4\pi}{3}\left((r+\mDelta)^3-r^3\right)$ of the
shell.
We then get
$\Vest_s^\mDelta(\bfx_i)\longrightarrow\Vt_s^\mDelta(\bfx_i)$ for a
large number
$\Nsh$ of (quasi-)random points. Consequently we compare
\begin{align}
  \label{eq:DRshell}
  \DR_{\text{shell}}(r) = \frac{1}{|\CW|N}\sum_{i=1}^N\Vest_r^\mDelta(\bfx_i)
\end{align}
with
$\DRt(r)$. This is not a pair-count, but we still have a double sum
over $N$ data points and now $\Nsh$ points in the shell.

\begin{figure}
  \begin{center}
    \includegraphics[width=0.38\textwidth]{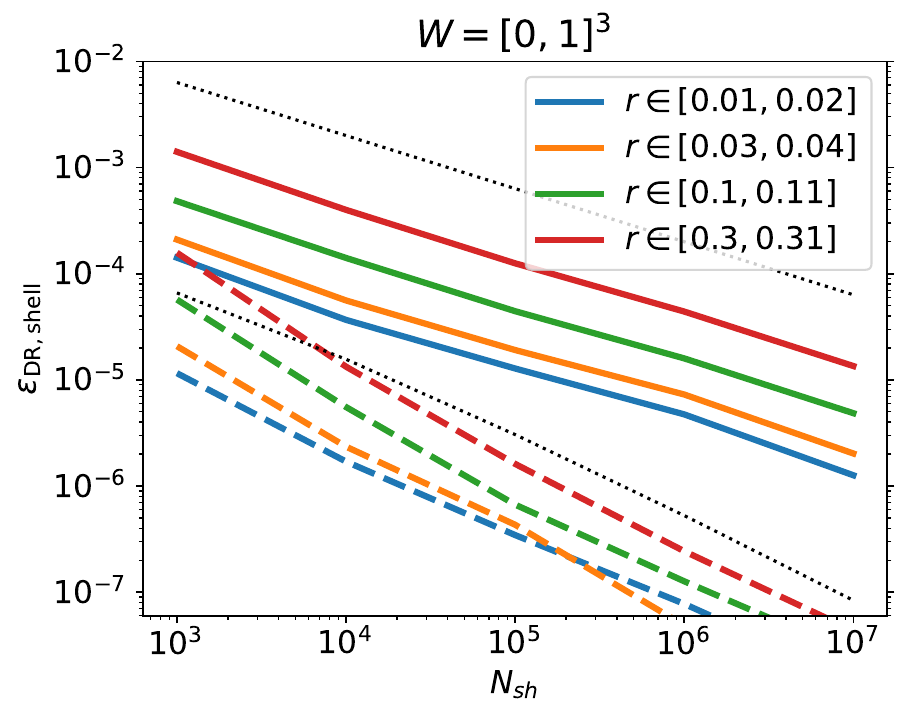}
    \includegraphics[width=0.38\textwidth]{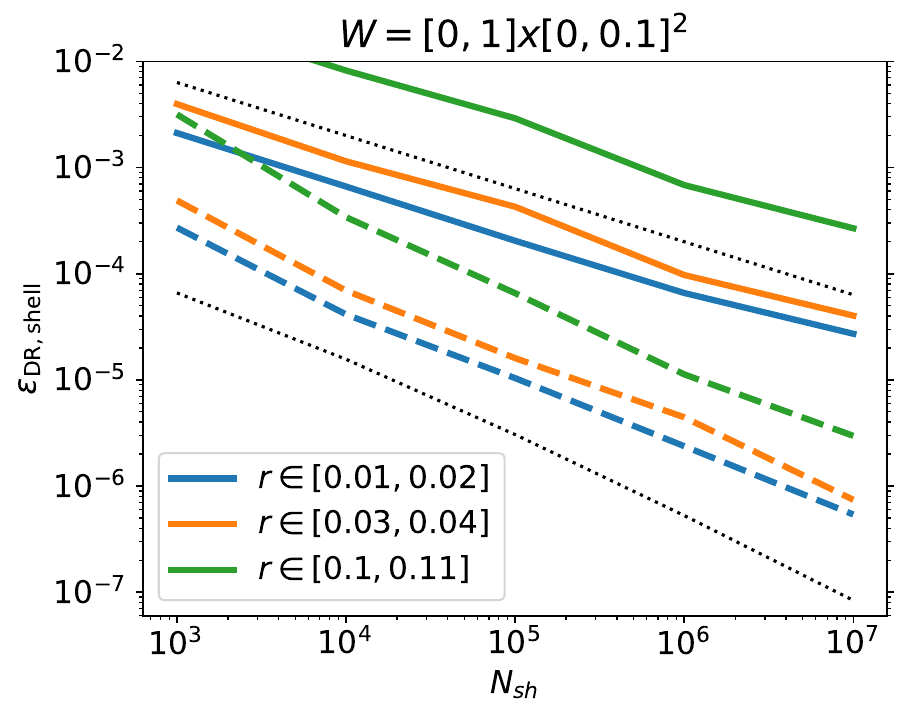}
  \end{center}
  \caption{ Relative error
    $\varepsilon_{\DR,\text{shell}}$ calculated for rectangular
    windows
    $\CW$ with a pure Monte~Carlo integration (solid lines) and with a
    randomised Halton sequence (dashed lines). The black dotted lines
    are proportional to $1/\sqrt{\Nsh}$ (upper) and $(\log
    \Nsh)^3/\Nsh$ (lower).}
  \label{fig:err_DRshell} 
\end{figure}

Figure\,\ref{fig:err_DRshell} shows
$\varepsilon_{\DR,\text{shell}}=|\DRt(r)-\DR_{\text{shell}}(r)|/\DRt(r)$.
A comparison with
$\DR$ in Fig.\,\ref{fig:epsDR} shows that using
$\DR_{\text{shell}}$ leads to a reduction of the error by almost two
orders of magnitude even for ordinary random points. Again this can be
improved by using low-discrepancy sequences; doing so allows us to
additionally gain at least another order of magnitude.

\section{Estimating $\xi$}
\label{sec:xi}

Now we join the improved pair-count estimates together and compare
results from the standard \citet{landy:bias} estimator to our new
estimator for the pair correlation function.  We use the simulated
galaxy clusters from the Magneticum simulation to illustrate the
behaviour of the different estimators
\citep{hirschmann:cosmological,ragagnin:web}, compare also
Sect.\,\ref{sec:DR}).  The 10429\,clusters are in a box with
side-length $325\hMpc$ and we use their real space positions.
The standard \citet{landy:bias} estimator was defined in
Eq.\,(\ref{eq:xiLS}):
\[
\xi_{\text{LS}}(r) = \frac{\DD(r)-2\DR(r)+\RR(r)}{\RR(r)} .
\]
First we use the same $\Nr$ random points to calculate $\RR$ and
$\DR$ (see Eqs.\,(\ref{eq:defDR}) and (\ref{eq:defRR})).
As an exact reference we have
\begin{align}
\Xi(r) & = \frac{\DD(r)-2\DRt(r)+\RRt(r)}{\RRt(r)} ,
\end{align}
with the $\RRt$ and $\DRt$ given in Eqs.\,(\ref{eq:expectRR}),
(\ref{eq:expectDR}), using the results from
Appendix\,\ref{sec:setcov-afrac-box} for a rectangular box.
In Sect.\,\ref{sec:RR} and \ref{sec:DR} we discuss alternative
possibilities to calculate the pair-counts $DR$ and $RR$.  We focus on
the following combination:
\begin{align}
  \label{eq:newestimator}
  \widetilde\xi_{\text{LS}}(r) & =
  \frac{\DD(r)-2\DR_{\text{shell}}(r)+\RR_{\text{2x3d}}(r)}{\RR_{\text{2x3d}}(r)} ,
\end{align}
which resembles the \citet{landy:bias} estimator, but now with improved
pair-count estimates.  We use a low-discrepancy sequence with
$N_{\text{shell}}$ 3D points to calculate $\DR_{\text{shell}}$ (see
Eq.\,\ref{eq:DRshell}), and another 6D low-discrepancy sequence with
$N_{\text{2x3d}}$ points to calculate $\RR_{\text{2x3d}}$ (see
Eq.\,\ref{eq:RR2x3d}).

\begin{figure}
  \begin{center}
    \includegraphics[width=0.38\textwidth]{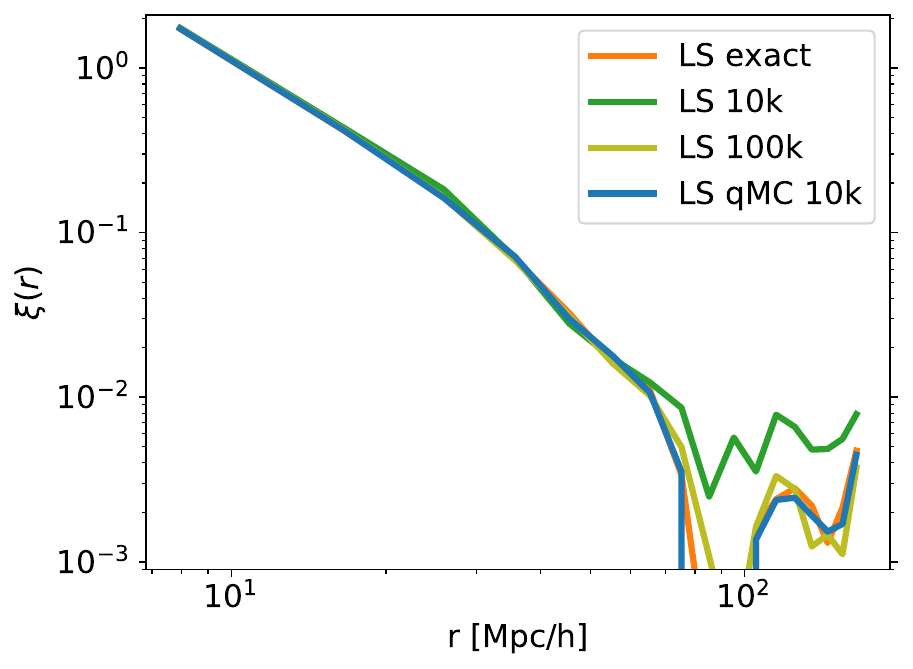}
    \includegraphics[width=0.38\textwidth]{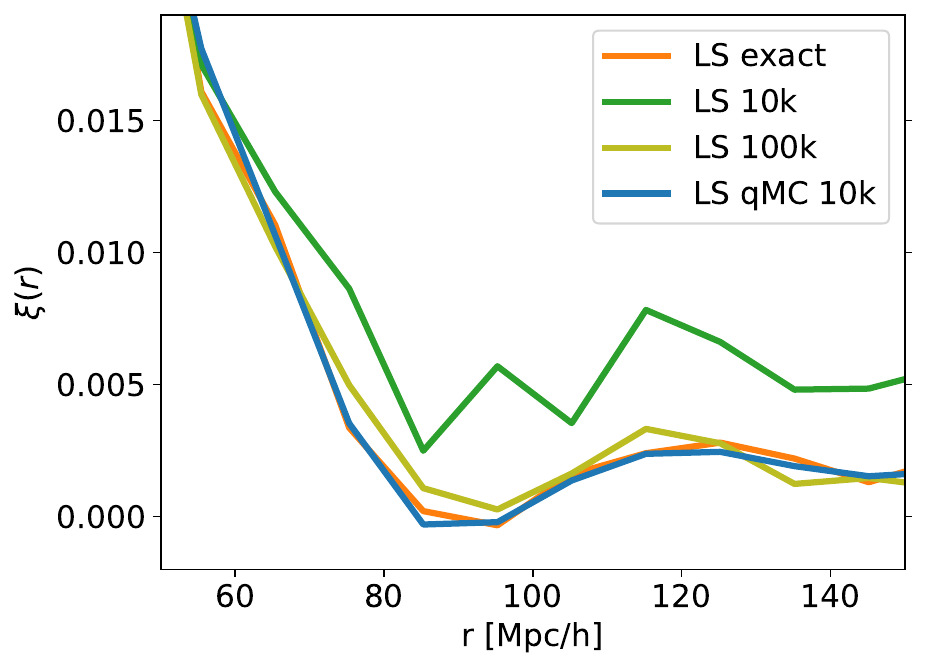}
  \end{center}
  \caption{ Two-point correlation function calculated for a
    simulated galaxy cluster sample. The exact $\Xi$ is compared to
    standard $\xi_{\text{LS}}$ with $\Nr=10^4$ and $\Nr=10^5$ random
    points, and to the new $\widetilde\xi_{\text{LS}}$ using
    $N_{\text{2x3d}}=10^4=\Nsh$ points from a randomised
    Halton sequence.}
  \label{fig:cluster_xi}
\end{figure}

As an illustrative example we compare these estimates in
Fig.\,\ref{fig:cluster_xi} using an insufficient number of points. As
expected, we observe that a standard \citet{landy:bias} estimator with
only $10^4$ random points shows deviations from the exact result, but
with $10^5$ random points the LS-estimator starts to follow the exact
result.  Visually, one can see that the estimator using the
low-discrepancy sequences achieves a higher accuracy already with
$N_{\text{2x3d}}=10^4=\Nsh$ points.

\begin{figure}
  \begin{center}
    \includegraphics[width=0.38\textwidth]{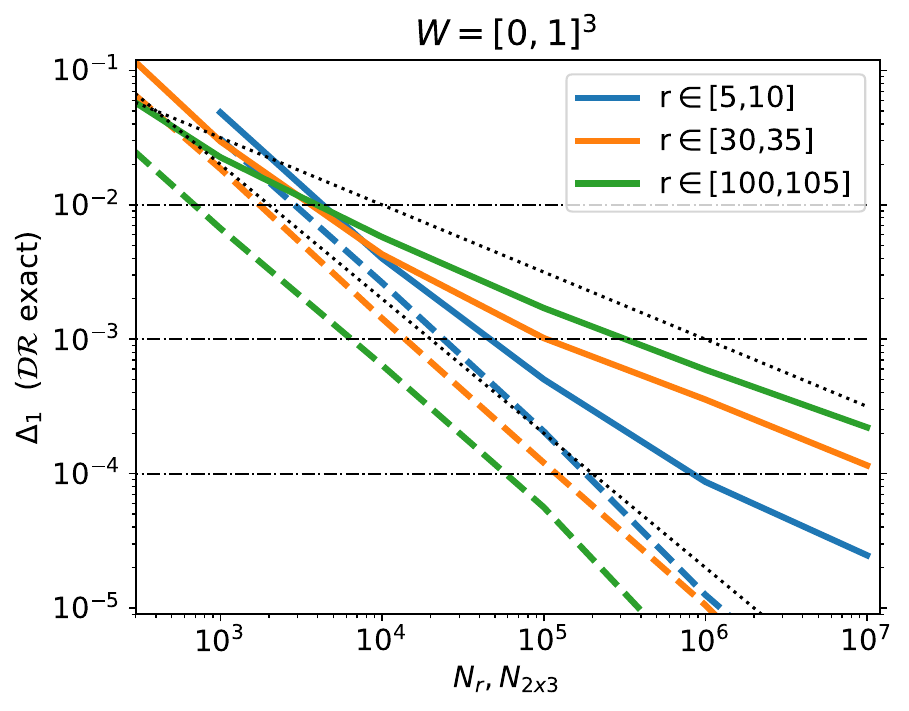}
    \includegraphics[width=0.38\textwidth]{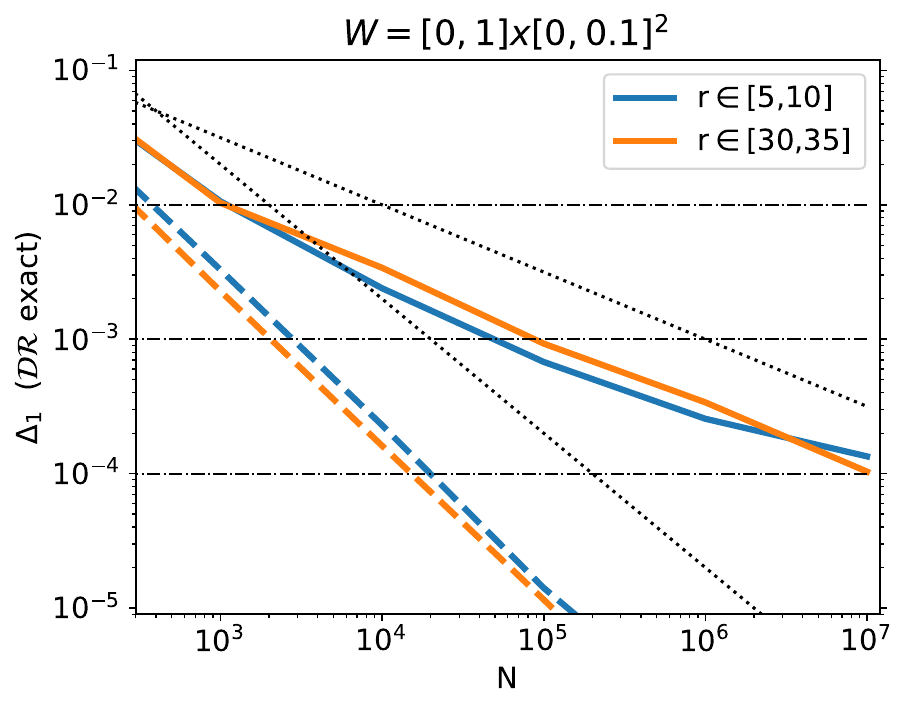}
  \end{center}
  \caption{ \label{fig:accuracy-halfexact1} Absolute error
    $\Delta_1$ is shown for the standard $\RR$
    (solid lines) and for $\RR_{\text{2x3d}}$ (dashed lines) pair-counts.
    The black dotted lines are proportional to $1/\sqrt{\Nr}$ (upper)
    and $1/N_{\text{2x3d}}$ (lower).}
\end{figure}

\begin{figure}
  \begin{center}
    \includegraphics[width=0.38\textwidth]{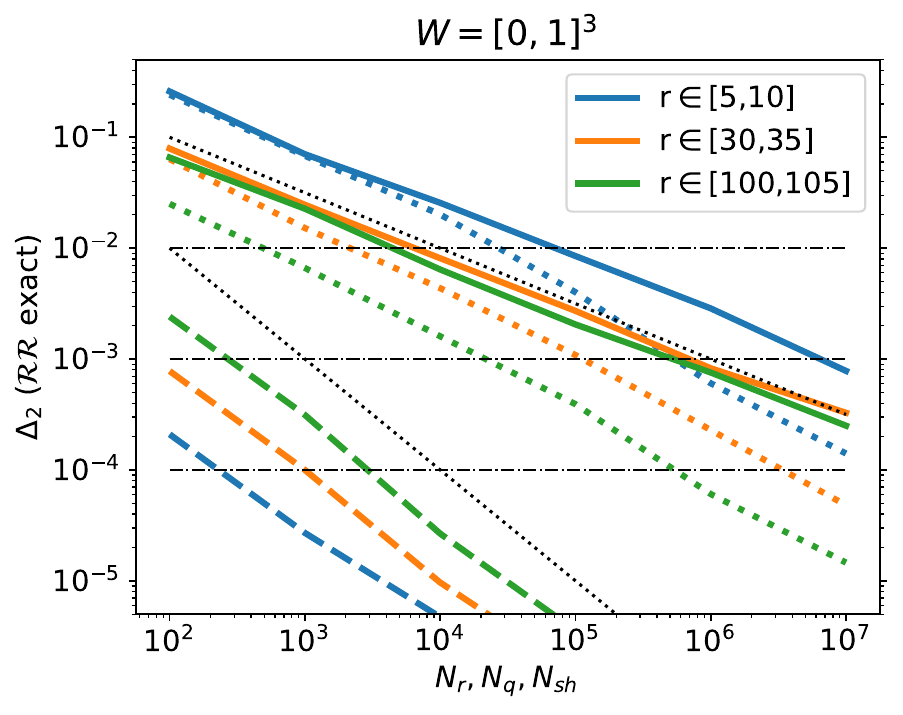}
    \includegraphics[width=0.38\textwidth]{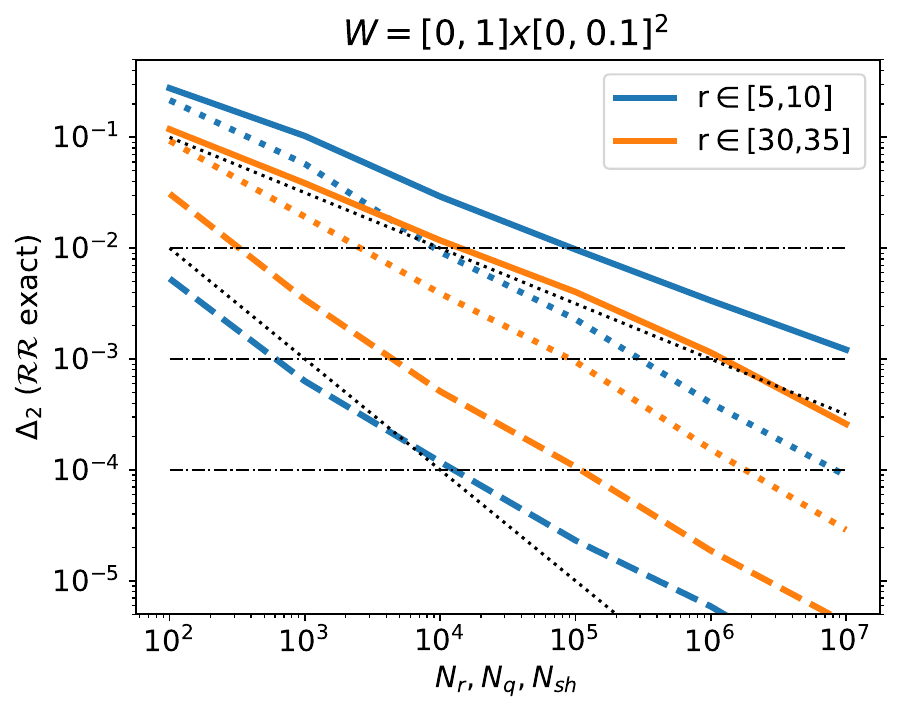}
  \end{center}
  \caption{  Absolute error
    $\Delta_2$ is shown for the standard $\DR$ (solid lines), the
    $\DR$ using a low-discrepancy sequence (dotted lines), and for 
    $\DR_{\text{shell}}$ also using a low-discrepancy sequence (dashed lines).
    The black dotted lines are proportional to $1/\sqrt{\Nr}$ (upper)
    and $1/\Nsh$ (lower).}
  \label{fig:accuracy-halfexact2}
\end{figure}

To make this quantitative and to disentangle the influence that these
different methods for calculating $\DR$ and $\RR$ have, we consider
the `half exact' estimates for $\xi(r):$
\begin{align}
  \Xi_{\RRt}(r) & = \frac{\DD(r)-2\DR(r)+\RRt(r)}{\RRt(r)}, \nonumber\\
  \Xi_{\DRt}(r) & = \frac{\DD(r)-2\DRt(r)+\RR(r)}{\RR(r)},
\end{align}
where we use the exact reference values $\RRt$ and $\DRt$ in turn.
We also calculate $\Xi_{\RRt}$ and $\Xi_{\DRt}$ using
$\DR_{\text{shell}}$ and $\RR_{\text{2x3d}}$ instead of $\DR$ and
$\RR$. For the comparison, we consider the absolute error
$\Delta_1(r)=|\Xi(r)-\Xi_{\DRt}(r)|$ and
$\Delta_2(r)=|\Xi(r)-\Xi_{\RRt}(r)|$.
In Fig.\,\ref{fig:accuracy-halfexact1} we see that the scaling
investigated in Sect.\,\ref{sec:RR} directly transfers to the scaling
of the error in the estimated $\xi$. For a fixed number of
(quasi-)random points, $\RR_{\text{2x3d}}$ is more accurate. The gain
in accuracy is considerable for intermediate and large radii. For
small radii and for a smaller number of (quasi-)random points, the
accuracy gain of $\RR_{\text{2x3d}}$ is reduced to a factor of two.
With the standard $\DR,$ the error $\Delta_1$ scales as $1/\sqrt{\Nr}$
(see Fig.\,\ref{fig:accuracy-halfexact2}). Already using a
low-discrepancy sequence instead of random points in $\DR$ leads to a
reduction in the error and the scaling starts to follow $1/\Nq$.
Compared to $\DR$, the $\DR_{\text{shell}}$ gives a significantly
smaller error, again showing a scaling proportional to $1/\Nsh$.  This
reduced error comes at a price: the run time of $\DR_{\text{shell}}$
is greater than the run time of $\DR$ for the same number of
(quasi-)random points (see Appendix\,\ref{sec:implementation}).

\subsection{Run time}
\label{sec:runtime}

Ultimately, we are not interested in the scaling of the error with the
number of (quasi-)random points. We want to obtain reliable results
{fast}.  The speed of the calculation depends on the hardware
platform, the algorithms, details of the implementation, and
further effects (see Appendix\,\ref{sec:implementation}).
Nevertheless, we prepared some examples and investigate what one can
expect as a speedup when using the new estimator.  In a given window,
and for a given radius interval, we look at
Figs.\,\ref{fig:accuracy-halfexact1} and \ref{fig:accuracy-halfexact2}
to read off how many (quasi-)random points we approximately need to
achieve a targeted absolute error. In most applications of the
standard estimator $\xi_{\text{LS}}$, the same random point set is used
with $\RR$ and $\DR$. However, there is some room for
optimisation. As we tune $\RR_{\text{2x3d}}$ and
$\DR_{\text{shell}}$ in the new estimator separately, we also choose
different sizes of the random samples in $\RR$ and $\DR$ to optimise
the overall run time.
With some provisional runs, we check that we achieve the targeted
accuracy $\Delta$ within 10\%; sometimes we had to adjust the
numbers. Table\,\ref{tab:cases}  lists the final parameters we
use. For case~III, we use $\Nr=10^7$ and achieve a
$\Delta\approx5\cdot10^{-4}$ for the standard estimator, and therefore the
run times for case~III may serve only as a lower bound.
\begin{table*}
  \caption{\label{tab:cases} Parameters and results of the run-time comparison.
    We abbreviate the voluminous window as
    $W_A$ and the thin window as $W_B$. The radius
    intervals are $R_{100}=[100,105]\,\hMpc$, $R_{30}=[30,35]\,\hMpc$,
    $R_{5}=[5,10]\,\hMpc$.  The targeted absolute accuracy is $\Delta$
    (compare the horizontal lines in
    Figs.\,\ref{fig:accuracy-halfexact1} and
    \,\ref{fig:accuracy-halfexact2}). 
    The run times for $\DD$ and so on are given in seconds.  }
\centering
\begin{tabular}{l || c|c|c || l|l|l|l || l|l|l|l|l  }
  \hline\hline
  \phantom{\LARGE I}& $r\in $ & $W$ & $\Delta$
  & $\Nr$ (RR) & $N_{\text{2x3d}}$ & $\Nr$ (DR) & $\Nsh$
  &  $T_{DD}$ & $T_{RR}$ & $T_{RR,\text{2x3d}}$ & $T_{DR}$ & $T_{DR,\text{shell}}$ \\
  \hline
  I\phantom{\Large I}  & $R_{100}$ & $W_A$ & $10^{-2}$
  & $7\cdot10^3$ & $8\cdot10^2$ & $7\cdot10^3$ & $7\cdot10$
  & $1\cdot10^{-1}$ &  $6\cdot10^{-2}$ & $5\cdot10^{-3}$ & $1\cdot10^{-1}$ & $1\cdot10^{-1}$ \\
  II & $R_{100}$ & $W_A$  & $10^{-3}$
  & $6\cdot10^5$ & $8\cdot10^3$ & $6\cdot10^5$ & $6\cdot10^2$
  & $1\cdot10^{-1}$ & $2\cdot10^2$ & $1\cdot10^{-1}$ & $5$ & $2\cdot10^{-1}$\\
  III& $R_{100}$ & $W_A$  & $10^{-4}$
  & $1\cdot10^7$ & $7\cdot10^4$ & $1\cdot10^7$  & $4\cdot10^3$
  & $1\cdot10^{-1}$ & $4\cdot10^{4}$  &  $4$ & $8\cdot10^{1}$ &  $8\cdot10^{-1}$ \\
  \hline
  IV\phantom{\Large I} & $R_{30}$  & $W_A$  & $10^{-3}$
  & $3\cdot10^5$  & $2\cdot10^4$ & $1\cdot10^6$ & $2\cdot10^2$
  & $1\cdot10^{-1}$ & $4\cdot10^1$ & $3\cdot10^{-1}$ & $7$ & $1\cdot10^{-1}$\\
  V  & $R_{30}$  & $W_B$  & $10^{-3}$
  & $3\cdot10^5$ & $4\cdot10^3$ & $3\cdot10^6$ & $6\cdot10^3$
  &  $2\cdot10^{-3}$ &  $5\cdot10^1$ & $4\cdot10^{-2}$ & $3\cdot10^{-1}$ & $9\cdot10^{-3}$\\
  \hline
  VI\phantom{\Large I} & $R_{5}$  & $W_A$  & $10^{-2}$
  & $6\cdot10^3$ & $3\cdot10^3$ & $1\cdot10^5$ & $5\cdot10$
  & $1\cdot10^{-1}$ & $4\cdot10^{-2}$ & $2\cdot10^{-2}$ & $6\cdot10^{-1}$  & $1\cdot10^{-1}$\\
  VII & $R_{5}$  & $W_A$  & $10^{-3}$
  & $6\cdot10^4$ &  $3\cdot10^4$ & $7\cdot10^6$ & $4\cdot10$
  & $1\cdot10^{-1}$ & $2$ & $7\cdot10^{-1}$ & $6\cdot10^1$ & $1\cdot10^{-1}$ \\
  \hline
  VIII\phantom{\Large I} & $R_{5}$  & $W_B$  & $10^{-2}$
  & $4\cdot10^3$ & $9\cdot10^2$ & $1\cdot10^5$ & $4\cdot10$
  & $7\cdot10^{-4}$ & $3\cdot10^{-2}$ & $6\cdot10^{-3}$ & $6\cdot10^{-3}$ & $1\cdot10^{-3}$ \\
  IX  & $R_{5}$  & $W_B$  & $10^{-3}$
  & $8\cdot10^4$ & $3\cdot10^3$ & $2\cdot10^7$ & $5\cdot10^3$
  & $1\cdot10^{-3}$ & $3$ & $3\cdot10^{-2}$ & $2$ & $1\cdot10^{-2}$ \\
  \hline
\end{tabular}
\end{table*}
The comparison of the run times was conducted on a small workstation
(for details see Appendix\,\ref{sec:implementation}). We estimate the
run times from 100 runs, for the longer running calculations we use 20
runs (in case~III 5~runs).
As the targeted accuracy is matched at the 10\% level and
this comparison did not run on an isolated machine, we can only give one
digit for the run times. See Appendix\,\ref{sec:implementation} for
further comments on the run time calculations.

The run times of the pair-counts are shown in Table\,\ref{tab:cases}
for different windows, radius ranges, and for different targeted
accuracies. From Appendix\,\ref{sec:implementation}, we know that for
a fixed number of (quasi-)random points, the standard pair-counts
$\RR$ and $\DR$ are always faster than the corresponding
$\RR_{\text{2x3d}}$ and $\DR_{\text{shell}}$. However, we choose the
number of (quasi-)random points by requiring a given accuracy. In that
case, we need fewer points for the new pair-counts, which in turn
lowers their run times.  Now we compare $\RR$ and $\RR_{\text{2x3d}}$
as well as $\DR$ and $\DR_{\text{shell}}$ at fixed accuracy, and find
that the run times of the new pair-counts are smaller in all the
considered cases.
We also see from Table\,\ref{tab:cases} that the relative
significance of the $\RR$, $\RR_{\text{2x3d}}$ and $\DR$,
$\DR_{\text{shell}}$ parts depends on the considered case. For a
targeted high accuracy (small $\Delta$), the run times are dominated by
the $\RR$ and $\RR_{\text{2x3d}}$ part, respectively. For small radii
and for lower accuracy, the $\DR$ and $\DR_{\text{shell}}$ parts
dominate the run time. The geometry of the window further
differentiates this picture.
In Table\,\ref{tab:summary}, we sum up the parts and show the run times
of $\xi_\text{LS}$ and the new $\widetilde{\xi}_\text{LS}$.  In all
the cases considered, we see a reduction in the run time using the new
estimator.
For high accuracy and for large and intermediate radii, the effect is
considerable. Using the new estimator we expect a reduction in
run time by a factor of between several hundred and almost $10^4$.
For small radii and for less demanding accuracy goals, our method
performs at least as well as the standard estimator and we may expect
a reduction by up to an order of magnitude. The timing results for
$\DD$, $\RR$, $\RR_{\text{2x3d}}$ and $\DR$ are independent of the
number of radius intervals, because we can sort the counts into the
radius bins.  However, we need to repeat the $\DR_{\text{shell}}$
calculation for each radius bin.
\begin{table}
  \caption{\label{tab:summary} We sum the parts and compare the
    run times for the standard $\xi_\text{LS}$ and the new
    $\widetilde{\xi}_\text{LS}$ in all the cases from
    Table\,\ref{tab:cases}. Times are in seconds.}
\centering
\begin{tabular}{l |l|l|r}
  \hline\hline
  \phantom{\LARGE I} & $T(\xi_\text{LS})$ &  $T(\widetilde{\xi}_\text{LS})$ & ratio\\
  \hline
  I\phantom{\Large I} & $3\cdot10^{-1}$ & $2\cdot10^{-1}$ & 1.5 \\
  II   & $2\cdot10^2$ & $5\cdot10^{-1}$ & 400 \\
  III  & $4\cdot10^4$ & $5$ & 8000\\
  \hline
  IV \phantom{\Large I}  & $4\cdot10^1$  & $5\cdot10^{-1}$ & 80 \\
  V    & $5\cdot10^1$  & $5\cdot10^{-2}$ & 1000\\
  \hline
  VI \phantom{\Large I}  & $8\cdot10^{-1}$ & $3\cdot10^{-1}$ & 2.7 \\
  VII  & $6\cdot10^1$ & $9\cdot10^{-1}$ & 67 \\
  \hline
  VIII\phantom{\Large I} & $3\cdot10^{-2}$ & $8\cdot10^{-3}$ & 3.75 \\
  IX   & $4$ & $4\cdot10^{-2}$ & 10 \\
  \hline
\end{tabular}
\end{table}

\subsection{A dense sample}
\label{sec:dense}

\begin{figure}
  \begin{center}
    \includegraphics[width=0.38\textwidth]{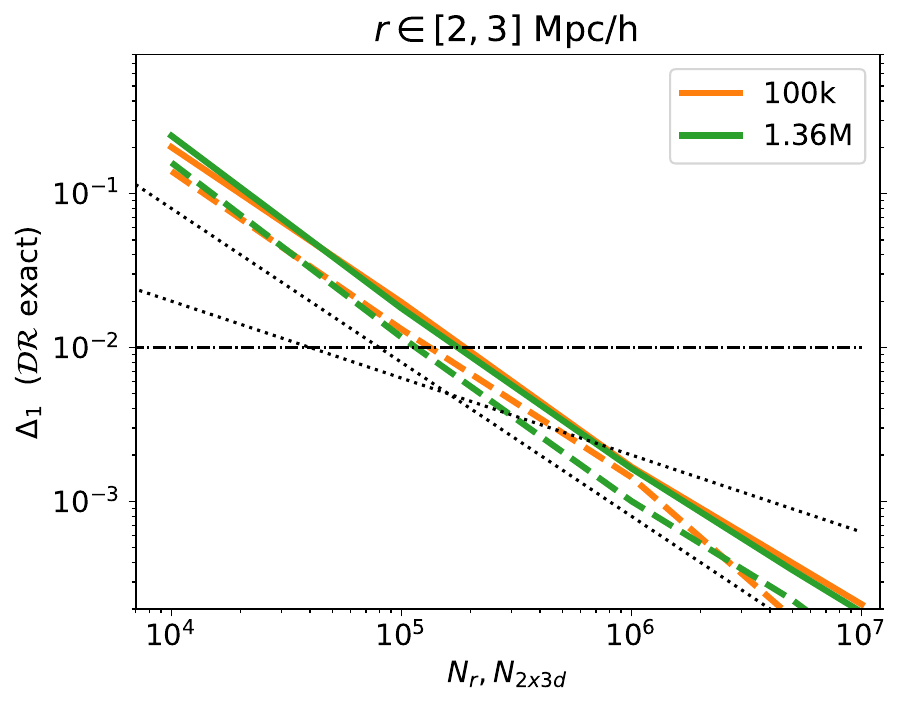}
    \includegraphics[width=0.38\textwidth]{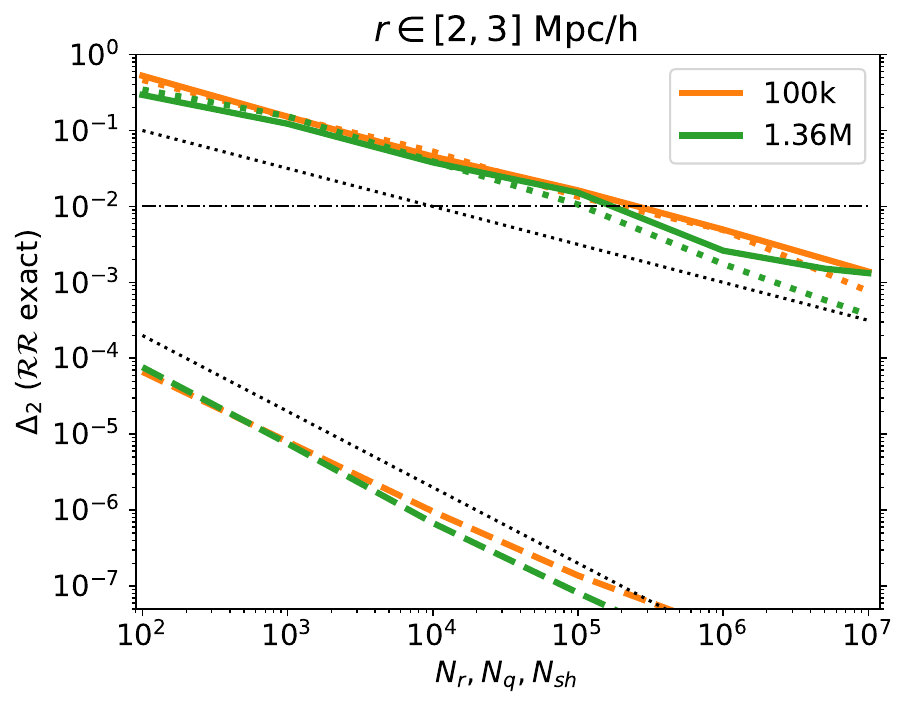}
  \end{center}
  \caption{ Errors for estimates of the correlation function at
    $r\in[2,3] \hMpc$ from 1.36~million galaxies and a subsample of
    100k galaxies in a box with side-length $325\hMpc$.
    Upper plot: Absolute error
    $\Delta_1$ shown for the standard $\RR$
    (solid lines) and for $\RR_{\text{2x3d}}$ (dashed lines) pair-counts.
    Lower plot: Absolute error
    $\Delta_2$ is shown for the standard $\DR$ (solid lines), the
    $\DR$ using a low-discrepancy sequence (dotted lines), and for 
    $\DR_{\text{shell}}$ also using a low-discrepancy sequence (dashed lines).
    The black dotted lines are proportional to $1/\sqrt{\Nr}$ (upper)
    and $1/N_{\text{2x3d}}$ (lower).}
  \label{fig:accuracy-halfexact-dense}
\end{figure}

Case~VI from Table\,\ref{tab:cases} shows that for small radii and
lower accuracy goals, the run time is dominated by the $\DR$ or the
$\DR_{\text{shell}}$ calculations. However, $\DR$ and $\DR_{\text{shell}}$
show very different run times depending on the number of data points
(see Appendix\,\ref{sec:implementation}). Hence per se it is not clear
which of the methods is preferable in dense samples for small radii
and relaxed accuracy goals.  We cannot check this with the cluster
sample with $N=10429$ data points only.
Therefore, we use a sample of 1.36~million simulated
galaxies\footnote{We use the simulated galaxies from the snapshot
  \texttt{Box2/hr, snap\_136, z=0.066340191} downloaded from
  \url{http://www.magneticum.org/data.html\#FULL_CATALOUGES}.}  from
the Magneticum simulation (\citealt{hirschmann:cosmological}; \citealt{ragagnin:web})
in the box with side-length $325\hMpc$.
In Fig.\,\ref{fig:accuracy-halfexact-dense}, we see that the errors of
$\RR$ and $\RR_{\text{2x3d}}$ both scale proportionally to $1/\Nr$
(see the discussion in Sect.\,\ref{sec:RR}).  For a fixed number of
(quasi-)random points, the $\RR_{\text{2x3d}}$ is about twice as
accurate as the $\RR$ for $r\in[2,3]\,\hMpc$ in this dense
sample. However, the accuracies of $\DR$ and $\DR_{\text{shell}}$
differ drastically. Using the same number of points, accuracy is
increased when using $\DR_{\text{shell}}$ instead of $\DR$ by at least
a factor of $10^3$.
We can understand this by looking at eq.\,(\ref{eq:DRshell}). In
calculating $\DR_{\text{shell}}$, we average the shell volume inside
the window over all the points. With so many points in this dense
sample, only a rough estimate of the shell volume for each individual
point or galaxy is necessary to obtain a $\DR_{\text{shell}}$ of
sufficient accuracy.

We proceed as in Sect.\ref{sec:runtime}, choose a targeted
accuracy $\Delta=10^{-2}$, and determine the required number of points
$\Nr$, $N_{\text{2x3d}}$, and $\Nsh$ in the pair-counts as
summarised in Table\,\ref{tab:cases2}.
Extrapolating from Fig.\,\ref{fig:accuracy-halfexact-dense}, we could
have guessed that the number $\Nsh$ of points, as used in
$\DR_{\text{shell}}$, is less than one. Instead of subsampling the
galaxies, we simply use $\Nsh=1$, and therefore we are overestimating
the required run time $T_{DR,\text{shell}}$.  It is also interesting to
observe that in case~XI we only need $\Nr\ll N$ random points for the
pair-counts in the standard estimators to achieve the targeted
accuracy.

From Table\,\ref{tab:cases2} we see that for the subsampled galaxy
distribution (case~X, 100k~points), the pair-counts $\RR$ and $\DR$
have a comparable contribution to the run time for the standard
estimator, whereas for the new estimator, the $\RR_{\text{2x3d}}$ still
provides the major contribution to the run time.
For the full galaxy sample with 1.36~million galaxies, the dominant
contribution is from $\DD$ followed by $\DR$ for the standard
estimator. For the new estimator, both $\RR_{\text{2x3d}}$ and
$\DR_{\text{shell}}$ contribute similarly to the run time.
In both cases, the new estimator is still faster than the standard
estimator, although only marginally in the full sample~XI.
\begin{table*}
  \caption{\label{tab:cases2} Parameters and results of a run time
    comparison similar to Table\,\ref{tab:cases} but now for a sample
    with $N=1.36$~million galaxies~(XI) and a random subsample with
    $N=10^5$ galaxies~(X) in a cube with side length $325\hMpc$.  We
    use the radius interval $[2,3]\,\hMpc$ and the targeted accuracy
    is $\Delta=10^{-2}$.  The run times for $\DD$ and so on are given in
    seconds.  }
\centering
\begin{tabular}{l || c || l|l|l|l || l|l|l|l|l  || l|l|c}
  \hline\hline
  \phantom{\LARGE I} & $N$ 
  & $\Nr$ (RR) & $N_{\text{2x3d}}$ & $\Nr$ (DR) & $\Nsh$
  &  $T_{DD}$ & $T_{RR}$ & $T_{RR,\text{2x3d}}$ & $T_{DR}$ & $T_{DR,\text{shell}}$
  & $T(\xi_\text{LS})$ &  $T(\widetilde{\xi}_\text{LS})$ & ratio\\
  \hline
  X \phantom{\LARGE I} & $10^5$ 
  & $3\cdot10^5$  & $1.5\cdot10^5$  & $4\cdot10^5$  & $1$ 
  & 4 & $3\cdot10^1$ & $1\cdot10^1$ & $3\cdot10^1$ & $9\cdot10^{-1}$
  & $6\cdot10^1$  & $2\cdot10^1$  & 3 \\  
  XI \phantom{\LARGE I} &  all 
  & $3\cdot10^5$ & $1.5\cdot10^5$  & $3\cdot10^5$  & $1$ 
  & $7\cdot10^2$ & $3\cdot10^1$ & $1\cdot10^1$ & $3\cdot10^2$ & $1\cdot10^1$
  & $1\cdot10^3$ & $7\cdot10^2$ & 1.4\\  
  \hline
\end{tabular}
\end{table*}
These timing results should be considered with some reservation. As
illustrated in Appendix\,\ref{sec:implementation}, a tree-based method
can speedup the $\DD$, $\RR$, $\DR$, and $\RR_{\text{2x3d}}$ for small
radii significantly. For $\DR_{\text{shell}}$, such a method for
increasing the speed is currently unknown. This certainly affects the
relative contribution of $\DR$ versus $\DR_{\text{shell}}$ and
depending on the situation, $\DR$ using a tree-based method can be
faster than $\DR_{\text{shell}}$.

\section{Incomplete sampling}
\label{sec:incomplete}

The random points in $\RR$ and $\DR$ are not only used to correct for
finite-size effects but also to correct for incomplete sampling.
Here, we follow the ideas presented in \citet{baddeley:nonandsemi} and
\citet{shaw:globally} and adapt them for the pair-counts. In this way,
we can still use the new improved estimators and also correct for
incomplete sampling.

Typically the galaxy distribution is observed incompletely.  For
example, unobserved regions around bright stars are masked. This leads
to holes in the observational window $\CW$ and can be dealt with using
the methods already described.  However, a partial sampling of the
galaxy distribution in crowded fields leaves us with an inhomogeneous
selection of the galaxy sample. Also, in magnitude-limited samples, we
have a systematic selection of the galaxies, depending on the distance
from us.
We model this by $p(\bfx)\in(0,1]$, the probability of
including or observing a galaxy at position $\bfx$.  We assume that this
$p(\bfx)$ is statistically independent of any other point in the
galaxy distribution.
For example, in a crowded field, only a fraction of the galaxies are
randomly targeted for spectroscopic follow up. For a galaxy in this
field, we have a $p(\bfx)$ equal to the fraction of targeted galaxies
in the field.
In a magnitude-limited sample, we observe all galaxies down to a
limiting brightness $l_{\text{lim}}$. The luminosity of a galaxy
$L(d(\bfx),l)$ at position $\bfx$ can be determined from the
brightness $l>l_{\text{lim}}$ and its luminosity distance $d(\bfx)$
from our galaxy (we neglect absorption for simplicity). Therefore, at
a distance of $d$ we only include galaxies with luminosity
$L>L(d,l_{\text{lim}})$ in our catalogue.  Now consider $F(L),$ the
distribution function of the absolute luminosities of all the galaxies
(i.e.\ the normalised cumulative luminosity function).  The fraction
of galaxies included at a distance $d(\bfx)$ is then
$p(\bfx)=1-F(L(d(\bfx),l_{\text{lim}}))$. We therefore need a good
model for the luminosity function.
The exclusion of a galaxy due to fibre collision cannot be modelled
with the independent thinning (see below), and the following approach
may only serve as an approximation.  Further selection and sampling
effects are discussed in \cite{ross:clustering}.

To investigate the effects of incomplete sampling, we assume an
unobserved homogeneous and isotropic galaxy distribution $G$ with
number density $\varrho_G$ and two-point correlation function
$\xi_G(r)$. Our goal is to estimate this $\xi_G(r)$.  The
inhomogeneous sampling, described by $p(\bfx)$, leads to an observed
inhomogeneous galaxy catalogue and is modelled in the following way:
\begin{equation}
  \label{eq:thinnedprocess}
  \textsf{D}'=\{\bfx_i\in G |  \bfx_i\in \CW \text{ and } u_i\le p(\bfx_i)\},
\end{equation}
where $u_i$ are independent random variables that are specifically
independent of the points and uniformly distributed on $[0,1]$.  This
closely follows the construction of an inhomogeneous Markov point
process by independent thinning as discussed by
\citet{baddeley:nonandsemi}.
The observed inhomogeneous point set $\{\bfx_i'\}_{i=1}^N$ is
considered a realisation of $\textsf{D}'$.  With the number density
$\varrho_G$ of the homogeneous galaxy distribution, the inhomogeneous
number density of this point process is
$\varrho'(\bfx)=p(\bfx)\,\varrho_G$.
The two-point density, the probability of observing a point at $\bfx$
and $\bfy$, is
\begin{align}
  \varrho_2'(\bfx,\bfy) = \varrho'(\bfx)\varrho'(\bfy)
  \left(1 + \xi'(|\bfx-\bfy|)\right). 
\end{align}
This is well defined for a certain class of point process as defined
in \citet{baddeley:nonandsemi}.  As the thinning is assumed to be
independent from the points, we have
\begin{displaymath}
  \xi_G(|\bfx-\bfy|) = \xi'(|\bfx-\bfy|) 
  = \frac{\varrho_2'(\bfx,\bfy)}{ \varrho_G^2\ p(\bfx) p(\bfy)} - 1 ,
\end{displaymath}
for the two-point correlation function.

In full analogy to Sect.\,\ref{sec:stuffweneed}, we define the
`inhomogeneous' pair-counts $\DD'$.  The points $\{\bfx_i'\}_{i=1}^N$ are
the incompletely sampled galaxies with positions $\bfx_i'$ inside the
observation window $\CW$. Then
\begin{equation}
\label{eq:defDDprime}
\DD'(r) = \frac{1}{N^2} \sum_{i=1}^N\sum_{j=1, j\ne i}^N\
\mDelta k_r^\mDelta(|\bfx_i'-\bfx_j'|).
\end{equation}
To account for the effect of incomplete sampling, one often applies
the thinning as described in Eq.\,(\ref{eq:thinnedprocess}) to the
random points as well. In this way, our random point set
$\{\bfy_i'\}_{i=1}^\Nr$ is a realisation of an inhomogeneous Poisson
process with number density $\tfrac{\Nr}{N} \varrho'(\bfx)$.
The pair-counts involving these inhomogeneous random points are
\begin{equation}
  \label{eq:defRRprime}
  \RR'(r) = \frac{1}{\Nr^2}\sum_{i=1}^{\Nr} \sum_{j=1, j\ne i}^{\Nr}\
  \mDelta k_r^\mDelta(|\bfy_i'-\bfy_j'|),
\end{equation}
and 
\begin{equation}
  \label{eq:defDRprime}
  \DR'(r) =  \frac{1}{N \Nr}  \sum_{i=1}^N\ \sum_{j=1}^{\Nr}\
  \mDelta k_r^\mDelta(|\bfx_i'-\bfy_j'|) .
\end{equation}
Appendix\,\ref{sec:inhomogeneous} shows that the
\cite{peebles:statisticalIII} estimator using $\DD'$ and $\RR'$ is
`ratio unbiased'. For the \citet{landy:bias} estimator, we combine
the pair-counts $\DD'$, $\DR'$, and $\RR'$ with
\begin{align}
  \label{eq:xiLSprime}
\xi'_\text{LS}(r) &= \frac{\DD'(r) - 2\DR'(r) + \RR'(r)}{\RR'(r)} ,
\end{align}
in full analogy to Eq.\,(\ref{eq:xiLS}). Now we show how
we can calculate these inhomogeneous pair-counts using our new
approach based on quasi-Monte~Carlo methods. 
For this we make a detour and consider the expectation 
value of $\RR'$ and then $\DR'$.

For a large number of random points $\Nr$ from an inhomogeneous Poisson
process, we obtain (see Appendix\,\ref{sec:inhomogeneous}):
\begin{align}
  \label{eq:expectRRprime}
  \RR'(r) & \rightarrow\ 
  \RRt'(r) =  \frac{4\pi}{N^2} \int_{r}^{r+\mDelta}
  \overline{\Gamma_\CW}(s) \ s^2\rmd s,
\end{align}
where $ \overline{\Gamma_\CW}(s)$ is given in Eq.\,(\ref{eq:GammaW}),
and $\RRt'(r)$ is the expectation value of the pair-count $\RR'(r)$.
There is no longer a direct correspondence with geometric objects like
the set-covariance, however we can write $\RRt'$ as the following
integral (compare to Eq.\,(\ref{eq:RR6d})):
\begin{multline}
  \label{eq:RR6dprime}
  \RRt'(r) = \frac{1}{N^2} \int_{\mathbb{R}^3}\! \int_{\mathbb{R}^3}
  \mathbb{1}_\CW(\bfy_1)\mathbb{1}_\CW(\bfy_2)\,
  \varrho'(\bfy_1)\varrho'(\bfy_2)\ \times\\
  \times\ \mDelta k_r^\mDelta(|\bfy_1-\bfy_2|) \, \rmd\bfy_1\,\rmd\bfy_2.
\end{multline}
This suggests a (quasi-)Monte~Carlo approach similar to
Eq.\,(\ref{eq:RR2x3d}). We use homogeneously sampled point sets
$P_1,P_2$ as described in Sect.\,\ref{sec:RR}. Then
\begin{align}
  \label{eq:RR2x3dprime}
  \RR_{2x3d}'(r)
  &= \frac{1}{N^2}\sum_{i=1}^{\Nr}
    \sum_{j=1}^{\Nr}\ \varrho'(\bfy_i)\varrho'(\bfz_j)\
    \mDelta k_r^\mDelta(|\bfy_i-\bfz_j|),
\end{align}
is an estimate of $\RRt'(r)$ (we reiterate that the number density of the
inhomogeneous Poisson process is $\tfrac{\Nr}{N} \varrho'(\bfx)$). As
described in Sect.\,\ref{sec:RR}, the point sets $P_1,P_2$ can be
generated from a six-dimensional random sequence or six-dimensional
low-discrepancy sequence.

In Eq.\,(\ref{eq:defDRprime}) we use a thinned random point set to
calculate $\DR'$. Similarly, we could use a thinned low-discrepancy
sequence in Eq.\,(\ref{eq:defDRprime}), but in Sect.\,\ref{sec:DR} we
see that this first approach is only mildly successful. We therefore
follow the second approach by considering Eq.\,(\ref{eq:DRtprime}) and
find
\begin{align}
  \DR'(r) & \rightarrow\ \DRt'(r)
            = \frac{1}{N^2} \sum_{i=1}^N \Ut_r^\mDelta(\bfx_i),
\end{align}
where $\Ut_r^\mDelta(\bfx_i)$ is given in Eq.\,(\ref{eq:Urdelta}).
Similar to Eq.\,(\ref{eq:Vest}), we use $\Nsh$ points
$\{\bfy_i\}_{i=1}^{\Nsh}$ (quasi-)randomly distributed in the
shell
$S_r^\mDelta(\bfx_i)=\{\bfy\in\bbR^3 \,|\, s < |\bfy-\bfx_i|\le
s+\mDelta \}$ around $\bfx_i$ to estimate $\Ut_r^\mDelta(\bfx_i)$ by
\begin{align}
  \Uest_s^\mDelta(\bfx_i) &=  \frac{|S_r^\mDelta|}{\Nsh}
  \sum_{j=1}^{\Nsh}\mathbb{1}_\CW(\bfy_j)\ \varrho'(\bfy_j).
\end{align}
This results in a new estimate for $\DRt'(r)$:
\begin{align}
  \label{eq:DRshellprime}
  \DR_{\text{shell}}'(r) = \frac{1}{N^2}\sum_{i=1}^N\Uest_r^\mDelta(\bfx_i) .
\end{align}
Similar to Eq.\,(\ref{eq:newestimator}) we can use $\DD'$,
$\RR_{2x3d}'$, and $\DR_{\text{shell}}'$ to construct a
\citet{landy:bias}-type estimator $\widetilde\xi_{\text{LS}}'$ for
inhomogeneously sampled galaxy catalogues.

\citet{shaw:globally} discuss how to estimate the $\xi'(r)$ {and}
a non-parametric model for $\varrho'(\bfx)$ directly from the data. In
a typical application to galaxy catalogues, we often have a good model
for $\varrho'(\bfx)$ (or equivalently for $p(\bfx)$).  Parameters of this
model are fixed by the sampling strategy, and some are determined from
the galaxy distribution (e.g.\ from an estimate of the luminosity
function).

\section{Summary and Outlook}
\label{sec:summary}

First, we focussed on the scaling of the error in estimates of the
pair-counts $\RR$ and $\DR$ with the number of (quasi-)random points.
For the standard approach, with ordinary random numbers, we confirm
the expected slow shrinking of the error proportional to $1/\sqrt{N}$.
A reformulation of the pair-counts makes a quasi-Monte~Carlo
integration possible. There we find that the error shrinks almost
proportionally to $1/\Nq$, where $\Nq$ is the number of points from a
low-discrepancy sequence.
This scaling of the error not only holds for bulky samples but is even
more pronounced in a thin sample with prevalent boundary effects.
We are therefore confident that our improved methods are also applicable
in more complicated sample geometries.
We combine these improved pair-counts into a new
\citet{landy:bias}-type estimator and compare with the standard
one. The new estimator inherits the favourable scaling proportional to
$1/\Nq$.

We can turn this observation around. For a fixed maximum error, we read
off how many points are necessary to stay below this error with both the
new and the standard estimator.
We then compare the run times of the estimators.  Depending on the
accuracy goal, the radius range considered, and the density of the
data, we observe a speedup by 50\%, and up to a factor of almost $10^4$
for our improved estimator.
More specifically, using $\RR_{\text{2x3d}}$ instead of the standard
$\RR$ will increases the efficiency (accuracy vs.\ run time) of the
estimator in any considered case. As tree-based methods are equally
applicable to $\RR$ and $\RR_{\text{2x3d}}$, we recommend using
$\RR_{\text{2x3d}}$ for any use case.
Also, in all the cases we considered, the $\DR_{\text{shell}}$ is
faster than the standard $\DR$ at fixed accuracy. However, an
unreserved recommendation is not possible. For the $\DR$ calculations,
a tree based method could be employed to speed up the calculations (we
did not use one in our calculations). Currently, no comparable method
for a speedup is known for $\DR_{\text{shell}}$. Therefore, in some
situations for fixed accuracy, a tree-based method could lead to
smaller run times in the $\DR$ calculations than in the
$\DR_{\text{shell}}$ calculations.  We expect this to be relevant for
voluminous samples, small radii, and lower accuracy goals.  A
combination of $\DD$, $\DR$, $\RR_{\text{2x3d}}$ similar to
eq.\,(\ref{eq:newestimator}) might therefore be useful. For large
radii, we recommend $\DR_{\text{shell}}$.

The random point sets are not only used for boundary corrections but
to correct for incomplete sampling as well. Typically, the selection and
sampling effects, as present in the galaxy distribution, are modelled
onto the random point set used in the pair-counts. We discuss how to
adapt this for the new improved estimator; we are using
the probability of observing the galaxies as a weight in the
calculation of the pair-counts.
However, this is only a first step. Using weighting schemes, one could
envision the construction of a minimum variance estimate for
inhomogeneous sampled galaxies, similarly to \citet{saunders:spatial},
\citet{feldman:power}, and \citet{colombi:effects}, but now for the
improved estimators using low-discrepancy sequences.

Our use of the randomised Halton sequence is similar to the use of
glass-like point sets by \citet{davila-kurban:improved}. Generating
glass-like point sets can be computationally challenging. In contrast,
a randomised Halton sequence is generated easily (see
Appendix\,\ref{sec:implementation}).  Halton sequences are probably
the simplest choice, but other low-discrepancy sequences could be used
as well \citep{lecuyer:recent}.
Interesting alternatives could be the so-called blue noise random sets,
which are used in computer graphics for efficiently sampling from
surfaces (see e.g.\ \citealt{heck:aliasingfree}).
The computational improvements like tree-based methods or optimised
memory layout are complimentary to our approach and can be used
similarly for the pair-counts with low-discrepancy sequences.  This
should lead to a further speedup.

\begin{acknowledgements}
  Many thanks to Adrian Baddeley for sharing code and the comments on
  the expressions for the area fraction.
  It is a pleasure to thank Klaus Dolag and Antonio Ragagnin for
  providing public access to the simulated galaxies and galaxy
  clusters from the Magneticum simulation.
  Also many thanks to the referee for his/her helpful comments and
  especially for the suggestion to include a more detailed discussion
  of the run times.
\end{acknowledgements}

\bibliographystyle{aa}
\bibliography{my}

\begin{thebibliography}{57}
\expandafter\ifx\csname natexlab\endcsname\relax\def\natexlab#1{#1}\fi

\bibitem[{{Abbott} {et~al.}(2022){Abbott}, {Aguena}, {Alarcon}, {Allam},
  {Alves}, {Amon}, {Andrade-Oliveira}, {Annis}, {Avila}, {Bacon}, {Baxter},
  {Bechtol}, {Becker}, {Bernstein}, {Bhargava}, {Birrer}, {Blazek},
  {Brandao-Souza}, {Bridle}, {Brooks}, {Buckley-Geer}, {Burke}, {Camacho},
  {Campos}, {Carnero Rosell}, {Carrasco Kind}, {Carretero}, {Castander},
  {Cawthon}, {Chang}, {Chen}, {Chen}, {Choi}, {Conselice}, {Cordero},
  {Costanzi}, {Crocce}, {da Costa}, {da Silva Pereira}, {Davis}, {Davis}, {De
  Vicente}, {DeRose}, {Desai}, {Di Valentino}, {Diehl}, {Dietrich}, {Dodelson},
  {Doel}, {Doux}, {Drlica-Wagner}, {Eckert}, {Eifler}, {Elsner}, {Elvin-Poole},
  {Everett}, {Evrard}, {Fang}, {Farahi}, {Fernandez}, {Ferrero}, {Fert{\'e}},
  {Fosalba}, {Friedrich}, {Frieman}, {Garc{\'\i}a-Bellido}, {Gatti},
  {Gaztanaga}, {Gerdes}, {Giannantonio}, {Giannini}, {Gruen}, {Gruendl},
  {Gschwend}, {Gutierrez}, {Harrison}, {Hartley}, {Herner}, {Hinton},
  {Hollowood}, {Honscheid}, {Hoyle}, {Huff}, {Huterer}, {Jain}, {James},
  {Jarvis}, {Jeffrey}, {Jeltema}, {Kovacs}, {Krause}, {Kron}, {Kuehn},
  {Kuropatkin}, {Lahav}, {Leget}, {Lemos}, {Liddle}, {Lidman}, {Lima}, {Lin},
  {MacCrann}, {Maia}, {Marshall}, {Martini}, {McCullough}, {Melchior},
  {Mena-Fern{\'a}ndez}, {Menanteau}, {Miquel}, {Mohr}, {Morgan}, {Muir},
  {Myles}, {Nadathur}, {Navarro-Alsina}, {Nichol}, {Ogando}, {Omori},
  {Palmese}, {Pandey}, {Park}, {Paz-Chinch{\'o}n}, {Petravick}, {Pieres},
  {Plazas Malag{\'o}n}, {Porredon}, {Prat}, {Raveri}, {Rodriguez-Monroy},
  {Rollins}, {Romer}, {Roodman}, {Rosenfeld}, {Ross}, {Rykoff}, {Samuroff},
  {S{\'a}nchez}, {Sanchez}, {Sanchez}, {Sanchez Cid}, {Scarpine}, {Schubnell},
  {Scolnic}, {Secco}, {Serrano}, {Sevilla-Noarbe}, {Sheldon}, {Shin}, {Smith},
  {Soares-Santos}, {Suchyta}, {Swanson}, {Tabbutt}, {Tarle}, {Thomas}, {To},
  {Troja}, {Troxel}, {Tucker}, {Tutusaus}, {Varga}, {Walker}, {Weaverdyck},
  {Wechsler}, {Weller}, {Yanny}, {Yin}, {Zhang}, {Zuntz}, \& {DES
  Collaboration}}]{abbott:des3}
{Abbott}, T.~M.~C., {Aguena}, M., {Alarcon}, A., {et~al.} 2022, Phys.\ Rev.\ D,
  105, 023520

\bibitem[{{Aghamousa} {et~al.}(2016){Aghamousa}, {Aguilar}, {Ahlen}, {Alam},
  {Allen}, {Allende Prieto}, {Annis}, {Bailey}, {Balland}, {Ballester},
  {Baltay}, {Beaufore}, {Bebek}, {Beers}, {Bell}, {Bernal}, {Besuner},
  {Beutler}, {Blake}, {Bleuler}, {Blomqvist}, {Blum}, {Bolton}, {Briceno},
  {Brooks}, {Brownstein}, {Buckley-Geer}, {Burden}, {Burtin}, {Busca}, {Cahn},
  {Cai}, {Cardiel-Sas}, {Carlberg}, {Carton}, {Casas}, {Castander},
  {Cervantes-Cota}, {Claybaugh}, {Close}, {Coker}, {Cole}, {Comparat},
  {Cooper}, {Cousinou}, {Crocce}, {Cuby}, {Cunningham}, {Davis}, {Dawson}, {de
  la Macorra}, {De Vicente}, {Delubac}, {Derwent}, {Dey}, {Dhungana}, {Ding},
  {Doel}, {Duan}, {Ealet}, {Edelstein}, {Eftekharzadeh}, {Eisenstein},
  {Elliott}, {Escoffier}, {Evatt}, {Fagrelius}, {Fan}, {Fanning}, {Farahi},
  {Farihi}, {Favole}, {Feng}, {Fernandez}, {Findlay}, {Finkbeiner},
  {Fitzpatrick}, {Flaugher}, {Flender}, {Font-Ribera}, {Forero-Romero},
  {Fosalba}, {Frenk}, {Fumagalli}, {Gaensicke}, {Gallo}, {Garcia-Bellido},
  {Gaztanaga}, {Pietro Gentile Fusillo}, {Gerard}, {Gershkovich},
  {Giannantonio}, {Gillet}, {Gonzalez-de-Rivera}, {Gonzalez-Perez}, {Gott},
  {Graur}, {Gutierrez}, {Guy}, {Habib}, {Heetderks}, {Heetderks}, {Heitmann},
  {Hellwing}, {Herrera}, {Ho}, {Holland}, {Honscheid}, {Huff}, {Hutchinson},
  {Huterer}, {Hwang}, {Illa Laguna}, {Ishikawa}, {Jacobs}, {Jeffrey},
  {Jelinsky}, {Jennings}, {Jiang}, {Jimenez}, {Johnson}, {Joyce}, {Jullo},
  {Juneau}, {Kama}, {Karcher}, {Karkar}, {Kehoe}, {Kennamer}, {Kent},
  {Kilbinger}, {Kim}, {Kirkby}, {Kisner}, {Kitanidis}, {Kneib}, {Koposov},
  {Kovacs}, {Koyama}, {Kremin}, {Kron}, {Kronig}, {Kueter-Young}, {Lacey},
  {Lafever}, {Lahav}, {Lambert}, {Lampton}, {Landriau}, {Lang}, {Lauer}, {Le
  Goff}, {Le Guillou}, {Le Van Suu}, {Lee}, {Lee}, {Leitner}, {Lesser}, {Levi},
  {L'Huillier}, {Li}, {Liang}, {Lin}, {Linder}, {Loebman}, {Luki{\'c}}, {Ma},
  {MacCrann}, {Magneville}, {Makarem}, {Manera}, {Manser}, {Marshall},
  {Martini}, {Massey}, {Matheson}, {McCauley}, {McDonald}, {McGreer},
  {Meisner}, {Metcalfe}, {Miller}, {Miquel}, {Moustakas}, {Myers}, {Naik},
  {Newman}, {Nichol}, {Nicola}, {Nicolati da Costa}, {Nie}, {Niz}, {Norberg},
  {Nord}, {Norman}, {Nugent}, {O'Brien}, {Oh}, {Olsen}, {Padilla},
  {Padmanabhan}, {Padmanabhan}, {Palanque-Delabrouille}, {Palmese},
  {Pappalardo}, {P{\^a}ris}, {Park}, {Patej}, {Peacock}, {Peiris}, {Peng},
  {Percival}, {Perruchot}, {Pieri}, {Pogge}, {Pollack}, {Poppett}, {Prada},
  {Prakash}, {Probst}, {Rabinowitz}, {Raichoor}, {Ree}, {Refregier}, {Regal},
  {Reid}, {Reil}, {Rezaie}, {Rockosi}, {Roe}, {Ronayette}, {Roodman}, {Ross},
  {Ross}, {Rossi}, {Rozo}, {Ruhlmann-Kleider}, {Rykoff}, {Sabiu}, {Samushia},
  {Sanchez}, {Sanchez}, {Schlegel}, {Schneider}, {Schubnell}, {Secroun},
  {Seljak}, {Seo}, {Serrano}, {Shafieloo}, {Shan}, {Sharples}, {Sholl},
  {Shourt}, {Silber}, {Silva}, {Sirk}, {Slosar}, {Smith}, {Smoot}, {Som},
  {Song}, {Sprayberry}, {Staten}, {Stefanik}, {Tarle}, {Sien Tie}, {Tinker},
  {Tojeiro}, {Valdes}, {Valenzuela}, {Valluri}, {Vargas-Magana}, {Verde},
  {Walker}, {Wang}, {Wang}, {Weaver}, {Weaverdyck}, {Wechsler}, {Weinberg},
  {White}, {Yang}, {Yeche}, {Zhang}, {Zhao}, {Zheng}, {Zhou}, {Zhou}, {Zhu},
  {Zou}, {Zu}, \& {DESI Collaboration}}]{aghamousa:desi}
{Aghamousa}, A., {Aguilar}, J., {Ahlen}, S., {et~al.} 2016, arXiv e-prints,
  arXiv:1611.00036

\bibitem[{{Alarcon} {et~al.}(2021){Alarcon}, {Gaztanaga}, {Eriksen}, {Baugh},
  {Cabayol}, {Casas}, {Carretero}, {Castander}, {De Vicente}, {Fernandez},
  {Garcia-Bellido}, {Hildebrandt}, {Hoekstra}, {Joachimi}, {Manzoni}, {Miquel},
  {Norberg}, {Padilla}, {Renard}, {Sanchez}, {Serrano}, {Sevilla-Noarbe},
  {Siudek}, \& {Tallada-Cresp{\'\i}}}]{alarcon:pau}
{Alarcon}, A., {Gaztanaga}, E., {Eriksen}, M., {et~al.} 2021, Mon.\ Not.\ Roy.\
  Astron.\ Soc., 501, 6103

\bibitem[{{Alonso}(2012)}]{alonso:cute}
{Alonso}, D. 2012, arXiv e-prints, arXiv:1210.1833

\bibitem[{{Amendola} {et~al.}(2018){Amendola}, {Appleby}, {Avgoustidis},
  {Bacon}, {Baker}, {Baldi}, {Bartolo}, {Blanchard}, {Bonvin}, {Borgani},
  {Branchini}, {Burrage}, {Camera}, {Carbone}, {Casarini}, {Cropper}, {de
  Rham}, {Dietrich}, {Di Porto}, {Durrer}, {Ealet}, {Ferreira}, {Finelli},
  {Garc{\'\i}a-Bellido}, {Giannantonio}, {Guzzo}, {Heavens}, {Heisenberg},
  {Heymans}, {Hoekstra}, {Hollenstein}, {Holmes}, {Hwang}, {Jahnke},
  {Kitching}, {Koivisto}, {Kunz}, {La Vacca}, {Linder}, {March}, {Marra},
  {Martins}, {Majerotto}, {Markovic}, {Marsh}, {Marulli}, {Massey}, {Mellier},
  {Montanari}, {Mota}, {Nunes}, {Percival}, {Pettorino}, {Porciani},
  {Quercellini}, {Read}, {Rinaldi}, {Sapone}, {Sawicki}, {Scaramella},
  {Skordis}, {Simpson}, {Taylor}, {Thomas}, {Trotta}, {Verde}, {Vernizzi},
  {Vollmer}, {Wang}, {Weller}, \& {Zlosnik}}]{amendola:euclid}
{Amendola}, L., {Appleby}, S., {Avgoustidis}, A., {et~al.} 2018, Living Reviews
  in Relativity, 21, 2

\bibitem[{Anderson {et~al.}(1999)Anderson, Bai, Bschof, Blackford, Demmel,
  Dongarra, {Du Croz}, Greenbaum, Hammarling, McKenney, \&
  Sorensen}]{anderson:lapack}
Anderson, E., Bai, Z., Bschof, C., {et~al.} 1999, {LAPACK} Users Guide, 3rd
  edn. (Philadelphia: SIAM)

\bibitem[{Baddeley \& Turner(2005)}]{baddeley:spatstat}
Baddeley, A. \& Turner, R. 2005, Journal of Statistical Software, 12, 1

\bibitem[{Baddeley {et~al.}(1993)Baddeley, Moyeed, Howard, \&
  Boyde}]{baddeley:3dpoint}
Baddeley, A.~J., Moyeed, R.~A., Howard, C.~V., \& Boyde, A. 1993, Appl.\
  Statist., 42, 641

\bibitem[{Baddeley {et~al.}(2000)Baddeley, Møller, \&
  Waagepetersen}]{baddeley:nonandsemi}
Baddeley, A.~J., Møller, J., \& Waagepetersen, R. 2000, Statistica
  Neerlandica, 54, 329

\bibitem[{{Bautista} {et~al.}(2021){Bautista}, {Paviot}, {Vargas Maga{\~n}a},
  {de la Torre}, {Fromenteau}, {Gil-Mar{\'\i}n}, {Ross}, {Burtin}, {Dawson},
  {Hou}, {Kneib}, {de Mattia}, {Percival}, {Rossi}, {Tojeiro}, {Zhao}, {Zhao},
  {Alam}, {Brownstein}, {Chapman}, {Choi}, {Chuang}, {Escoffier}, {de la
  Macorra}, {du Mas des Bourboux}, {Mohammad}, {Moon}, {M{\"u}ller},
  {Nadathur}, {Newman}, {Schneider}, {Seo}, \& {Wang}}]{bautista:complete}
{Bautista}, J.~E., {Paviot}, R., {Vargas Maga{\~n}a}, M., {et~al.} 2021, Mon.\
  Not.\ Roy.\ Astron.\ Soc., 500, 736

\bibitem[{{Breton} \& {de la Torre}(2021)}]{breton:fast}
{Breton}, M.-A. \& {de la Torre}, S. 2021, A\&A, 646, A40

\bibitem[{Colombi {et~al.}(1998)Colombi, Szapudi, \& Szalay}]{colombi:effects}
Colombi, S., Szapudi, I., \& Szalay, A.~S. 1998, Mon.\ Not.\ Roy.\ Astron.\
  Soc., 296, 253

\bibitem[{Dagum \& Menon(1998)}]{dagum:openmp}
Dagum, L. \& Menon, R. 1998, Computational Science \& Engineering, IEEE, 5, 46

\bibitem[{Daley \& Vere-Jones(2003)}]{daley:introductionI}
Daley, D.~J. \& Vere-Jones, D. 2003, An Introduction to the Theory of Point
  Processes (Berlin: Springer Verlag)

\bibitem[{{D{\'a}vila-Kurb{\'a}n} {et~al.}(2021){D{\'a}vila-Kurb{\'a}n},
  {S{\'a}nchez}, {Lares}, \& {Ruiz}}]{davila-kurban:improved}
{D{\'a}vila-Kurb{\'a}n}, F., {S{\'a}nchez}, A.~G., {Lares}, M., \& {Ruiz},
  A.~N. 2021, Mon.\ Not.\ Roy.\ Astron.\ Soc., 506, 4667

\bibitem[{Davis \& Peebles(1983)}]{davis:surveyV}
Davis, M. \& Peebles, P. J.~E. 1983, ApJ, 267, 465

\bibitem[{{Dawson} {et~al.}(2013){Dawson}, {Schlegel}, {Ahn}, {Anderson},
  {Aubourg}, {Bailey}, {Barkhouser}, {Bautista}, {Beifiori}, {Berlind},
  {Bhardwaj}, {Bizyaev}, {Blake}, {Blanton}, {Blomqvist}, {Bolton}, {Borde},
  {Bovy}, {Brandt}, {Brewington}, {Brinkmann}, {Brown}, {Brownstein}, {Bundy},
  {Busca}, {Carithers}, {Carnero}, {Carr}, {Chen}, {Comparat}, {Connolly},
  {Cope}, {Croft}, {Cuesta}, {da Costa}, {Davenport}, {Delubac}, {de Putter},
  {Dhital}, {Ealet}, {Ebelke}, {Eisenstein}, {Escoffier}, {Fan}, {Filiz Ak},
  {Finley}, {Font-Ribera}, {G{\'e}nova-Santos}, {Gunn}, {Guo}, {Haggard},
  {Hall}, {Hamilton}, {Harris}, {Harris}, {Ho}, {Hogg}, {Holder}, {Honscheid},
  {Huehnerhoff}, {Jordan}, {Jordan}, {Kauffmann}, {Kazin}, {Kirkby}, {Klaene},
  {Kneib}, {Le Goff}, {Lee}, {Long}, {Loomis}, {Lundgren}, {Lupton}, {Maia},
  {Makler}, {Malanushenko}, {Malanushenko}, {Mandelbaum}, {Manera}, {Maraston},
  {Margala}, {Masters}, {McBride}, {McDonald}, {McGreer}, {McMahon}, {Mena},
  {Miralda-Escud{\'e}}, {Montero-Dorta}, {Montesano}, {Muna}, {Myers},
  {Naugle}, {Nichol}, {Noterdaeme}, {Nuza}, {Olmstead}, {Oravetz}, {Oravetz},
  {Owen}, {Padmanabhan}, {Palanque-Delabrouille}, {Pan}, {Parejko},
  {P{\^a}ris}, {Percival}, {P{\'e}rez-Fournon}, {P{\'e}rez-R{\`a}fols},
  {Petitjean}, {Pfaffenberger}, {Pforr}, {Pieri}, {Prada}, {Price-Whelan},
  {Raddick}, {Rebolo}, {Rich}, {Richards}, {Rockosi}, {Roe}, {Ross}, {Ross},
  {Rossi}, {Rubi{\~n}o-Martin}, {Samushia}, {S{\'a}nchez}, {Sayres}, {Schmidt},
  {Schneider}, {Sc{\'o}ccola}, {Seo}, {Shelden}, {Sheldon}, {Shen}, {Shu},
  {Slosar}, {Smee}, {Snedden}, {Stauffer}, {Steele}, {Strauss}, {Streblyanska},
  {Suzuki}, {Swanson}, {Tal}, {Tanaka}, {Thomas}, {Tinker}, {Tojeiro},
  {Tremonti}, {Vargas Maga{\~n}a}, {Verde}, {Viel}, {Wake}, {Watson}, {Weaver},
  {Weinberg}, {Weiner}, {West}, {White}, {Wood-Vasey}, {Yeche}, {Zehavi},
  {Zhao}, \& {Zheng}}]{dawson:boss}
{Dawson}, K.~S., {Schlegel}, D.~J., {Ahn}, C.~P., {et~al.} 2013, AJ, 145, 10

\bibitem[{Demina {et~al.}(2018)Demina, Cheong, BenZvi, \&
  Hindrichs}]{demina:computationally}
Demina, R., Cheong, S., BenZvi, S., \& Hindrichs, O. 2018, Mon.\ Not.\ Roy.\
  Astron.\ Soc., 480, 49

\bibitem[{{Donoso}(2019)}]{donoso:gundam}
{Donoso}, E. 2019, Mon.\ Not.\ Roy.\ Astron.\ Soc., 487, 2824

\bibitem[{{Eisenstein} {et~al.}(2005){Eisenstein}, {Zehavi}, {Hogg},
  {Scoccimarro}, {Blanton}, {Nichol}, {Scranton}, {Seo}, {Tegmark}, {Zheng},
  {Anderson}, {Annis}, {Bahcall}, {Brinkmann}, {Burles}, {Castander},
  {Connolly}, {Csabai}, {Doi}, {Fukugita}, {Frieman}, {Glazebrook}, {Gunn},
  {Hendry}, {Hennessy}, {Ivezi{\'c}}, {Kent}, {Knapp}, {Lin}, {Loh}, {Lupton},
  {Margon}, {McKay}, {Meiksin}, {Munn}, {Pope}, {Richmond}, {Schlegel},
  {Schneider}, {Shimasaku}, {Stoughton}, {Strauss}, {SubbaRao}, {Szalay},
  {Szapudi}, {Tucker}, {Yanny}, \& {York}}]{eisenstein:baryon}
{Eisenstein}, D.~J., {Zehavi}, I., {Hogg}, D.~W., {et~al.} 2005, ApJ, 633, 560

\bibitem[{Feldman {et~al.}(1994)Feldman, Kaiser, \& Peacock}]{feldman:power}
Feldman, H.~A., Kaiser, N., \& Peacock, J.~A. 1994, ApJ, 426, 23

\bibitem[{Fiksel(1988)}]{fiksel:edge}
Fiksel, T. 1988, Statistics, 19, 67

\bibitem[{Halton(1960)}]{halton:efficiency}
Halton, J. 1960, Numer. Math., 2, 84

\bibitem[{Hamilton(1993)}]{hamilton:towards}
Hamilton, A. 1993, ApJ, 417, 19

\bibitem[{Harris {et~al.}(2020)Harris, Millman, van~der Walt, Gommers,
  Virtanen, Cournapeau, Wieser, Taylor, Berg, Smith, Kern, Picus, Hoyer, van
  Kerkwijk, Brett, Haldane, del R{\'{i}}o, Wiebe, Peterson,
  G{\'{e}}rard-Marchant, Sheppard, Reddy, Weckesser, Abbasi, Gohlke, \&
  Oliphant}]{harris:numpy}
Harris, C.~R., Millman, K.~J., van~der Walt, S.~J., {et~al.} 2020, Nature, 585,
  357

\bibitem[{He(2021)}]{he:fast}
He, C.-C. 2021, ApJ, 921, 59

\bibitem[{Heck {et~al.}(2013)Heck, Schlömer, \& Deussen}]{heck:aliasingfree}
Heck, D., Schlömer, T., \& Deussen, O. 2013, ACM Transactions on Graphics, 32
  (3), 1

\bibitem[{Hewett(1982)}]{hewett:estimation}
Hewett, P.~C. 1982, Mon.\ Not.\ Roy.\ Astron.\ Soc., 201, 867

\bibitem[{{Hirschmann} {et~al.}(2014){Hirschmann}, {Dolag}, {Saro}, {Bachmann},
  {Borgani}, \& {Burkert}}]{hirschmann:cosmological}
{Hirschmann}, M., {Dolag}, K., {Saro}, A., {et~al.} 2014, Mon.\ Not.\ Roy.\
  Astron.\ Soc., 442, 2304

\bibitem[{{Ivezi{\'c}} {et~al.}(2019){Ivezi{\'c}}, {Kahn}, {Tyson}, {Abel},
  {Acosta}, {Allsman}, {Alonso}, {AlSayyad}, {Anderson}, {Andrew}, {Angel},
  {Angeli}, {Ansari}, {Antilogus}, {Araujo}, {Armstrong}, {Arndt}, {Astier},
  {Aubourg}, {Auza}, {Axelrod}, {Bard}, {Barr}, {Barrau}, {Bartlett}, {Bauer},
  {Bauman}, {Baumont}, {Bechtol}, {Bechtol}, {Becker}, {Becla}, {Beldica},
  {Bellavia}, {Bianco}, {Biswas}, {Blanc}, {Blazek}, {Blandford}, {Bloom},
  {Bogart}, {Bond}, {Booth}, {Borgland}, {Borne}, {Bosch}, {Boutigny},
  {Brackett}, {Bradshaw}, {Brandt}, {Brown}, {Bullock}, {Burchat}, {Burke},
  {Cagnoli}, {Calabrese}, {Callahan}, {Callen}, {Carlin}, {Carlson},
  {Chandrasekharan}, {Charles-Emerson}, {Chesley}, {Cheu}, {Chiang}, {Chiang},
  {Chirino}, {Chow}, {Ciardi}, {Claver}, {Cohen-Tanugi}, {Cockrum}, {Coles},
  {Connolly}, {Cook}, {Cooray}, {Covey}, {Cribbs}, {Cui}, {Cutri}, {Daly},
  {Daniel}, {Daruich}, {Daubard}, {Daues}, {Dawson}, {Delgado}, {Dellapenna},
  {de Peyster}, {de Val-Borro}, {Digel}, {Doherty}, {Dubois},
  {Dubois-Felsmann}, {Durech}, {Economou}, {Eifler}, {Eracleous}, {Emmons},
  {Fausti Neto}, {Ferguson}, {Figueroa}, {Fisher-Levine}, {Focke}, {Foss},
  {Frank}, {Freemon}, {Gangler}, {Gawiser}, {Geary}, {Gee}, {Geha}, {Gessner},
  {Gibson}, {Gilmore}, {Glanzman}, {Glick}, {Goldina}, {Goldstein}, {Goodenow},
  {Graham}, {Gressler}, {Gris}, {Guy}, {Guyonnet}, {Haller}, {Harris},
  {Hascall}, {Haupt}, {Hernandez}, {Herrmann}, {Hileman}, {Hoblitt}, {Hodgson},
  {Hogan}, {Howard}, {Huang}, {Huffer}, {Ingraham}, {Innes}, {Jacoby}, {Jain},
  {Jammes}, {Jee}, {Jenness}, {Jernigan}, {Jevremovi{\'c}}, {Johns}, {Johnson},
  {Johnson}, {Jones}, {Juramy-Gilles}, {Juri{\'c}}, {Kalirai}, {Kallivayalil},
  {Kalmbach}, {Kantor}, {Karst}, {Kasliwal}, {Kelly}, {Kessler}, {Kinnison},
  {Kirkby}, {Knox}, {Kotov}, {Krabbendam}, {Krughoff}, {Kub{\'a}nek},
  {Kuczewski}, {Kulkarni}, {Ku}, {Kurita}, {Lage}, {Lambert}, {Lange},
  {Langton}, {Le Guillou}, {Levine}, {Liang}, {Lim}, {Lintott}, {Long},
  {Lopez}, {Lotz}, {Lupton}, {Lust}, {MacArthur}, {Mahabal}, {Mandelbaum},
  {Markiewicz}, {Marsh}, {Marshall}, {Marshall}, {May}, {McKercher}, {McQueen},
  {Meyers}, {Migliore}, {Miller}, {Mills}, {Miraval}, {Moeyens}, {Moolekamp},
  {Monet}, {Moniez}, {Monkewitz}, {Montgomery}, {Morrison}, {Mueller},
  {Muller}, {Mu{\~n}oz Arancibia}, {Neill}, {Newbry}, {Nief}, {Nomerotski},
  {Nordby}, {O'Connor}, {Oliver}, {Olivier}, {Olsen}, {O'Mullane}, {Ortiz},
  {Osier}, {Owen}, {Pain}, {Palecek}, {Parejko}, {Parsons}, {Pease},
  {Peterson}, {Peterson}, {Petravick}, {Libby Petrick}, {Petry},
  {Pierfederici}, {Pietrowicz}, {Pike}, {Pinto}, {Plante}, {Plate}, {Plutchak},
  {Price}, {Prouza}, {Radeka}, {Rajagopal}, {Rasmussen}, {Regnault}, {Reil},
  {Reiss}, {Reuter}, {Ridgway}, {Riot}, {Ritz}, {Robinson}, {Roby}, {Roodman},
  {Rosing}, {Roucelle}, {Rumore}, {Russo}, {Saha}, {Sassolas}, {Schalk},
  {Schellart}, {Schindler}, {Schmidt}, {Schneider}, {Schneider}, {Schoening},
  {Schumacher}, {Schwamb}, {Sebag}, {Selvy}, {Sembroski}, {Seppala}, {Serio},
  {Serrano}, {Shaw}, {Shipsey}, {Sick}, {Silvestri}, {Slater}, {Smith},
  {Smith}, {Sobhani}, {Soldahl}, {Storrie-Lombardi}, {Stover}, {Strauss},
  {Street}, {Stubbs}, {Sullivan}, {Sweeney}, {Swinbank}, {Szalay}, {Takacs},
  {Tether}, {Thaler}, {Thayer}, {Thomas}, {Thornton}, {Thukral}, {Tice},
  {Trilling}, {Turri}, {Van Berg}, {Vanden Berk}, {Vetter}, {Virieux},
  {Vucina}, {Wahl}, {Walkowicz}, {Walsh}, {Walter}, {Wang}, {Wang}, {Warner},
  {Wiecha}, {Willman}, {Winters}, {Wittman}, {Wolff}, {Wood-Vasey}, {Wu},
  {Xin}, {Yoachim}, \& {Zhan}}]{ivezic:lsst}
{Ivezi{\'c}}, {\v{Z}}., {Kahn}, S.~M., {Tyson}, J.~A., {et~al.} 2019, ApJ, 873,
  111

\bibitem[{Jakob {et~al.}(2017)Jakob, Rhinelander, \& Moldovan}]{jakob:pybind11}
Jakob, W., Rhinelander, J., \& Moldovan, D. 2017, pybind11 -- Seamless
  operability between C++11 and Python, https://github.com/pybind/pybind11

\bibitem[{{Keih{\"a}nen} {et~al.}(2019){Keih{\"a}nen}, {Kurki-Suonio},
  {Lindholm}, {Viitanen}, {Suur-Uski}, {Allevato}, {Branchini}, {Marulli},
  {Norberg}, {Tavagnacco}, {de la Torre}, {Valiviita}, {Viel}, {Bel},
  {Frailis}, \& {S{\'a}nchez}}]{keihaennen:estimating}
{Keih{\"a}nen}, E., {Kurki-Suonio}, H., {Lindholm}, V., {et~al.} 2019, A\&A,
  631, A73

\bibitem[{Kerscher(1999)}]{kerscher:twopoint}
Kerscher, M. 1999, Astron.~Astrophys., 343, 333

\bibitem[{Kerscher {et~al.}(2000)Kerscher, Szapudi, \&
  Szalay}]{kerscher:comparison}
Kerscher, M., Szapudi, I., \& Szalay, A. 2000, Ap.~J., 535, L13

\bibitem[{Landy \& Szalay(1993)}]{landy:bias}
Landy, S.~D. \& Szalay, A.~S. 1993, ApJ, 412, 64

\bibitem[{L'Ecuyer \& Lemieux(2002)}]{lecuyer:recent}
L'Ecuyer, P. \& Lemieux, C. 2002, Recent Advances in Randomized Quasi-Monte
  Carlo Methods (New York, NY: Springer US), 419--474

\bibitem[{{Moore} {et~al.}(2001){Moore}, {Connolly}, {Genovese}, {Gray},
  {Grone}, {Kanidoris}, {Nichol}, {Schneider}, {Szalay}, {Szapudi}, \&
  {Wasserman}}]{moore:fast}
{Moore}, A.~W., {Connolly}, A.~J., {Genovese}, C., {et~al.} 2001, in Mining the
  Sky, ed. A.~J. {Banday}, S.~{Zaroubi}, \& M.~{Bartelmann}, 71

\bibitem[{Neyman \& Scott(1958)}]{neyman:statistical}
Neyman, J. \& Scott, E.~L. 1958, J.\ R.\ Stat.\ Soc., 20, 1

\bibitem[{Niederreiter(1992)}]{niederreiter:random}
Niederreiter, H. 1992, Random Number Generation and Quasi-Monte Carlo Methods
  (Philadelphia, Pennsylvania: SIAM)

\bibitem[{Ohser(1983)}]{ohser:estimators}
Ohser, J. 1983, Math.\ Operationsforsch.\ u.\ Statist., Ser.\ Statist., 14, 63

\bibitem[{{Owen}(2017)}]{owen:randomized}
{Owen}, A.~B. 2017, arXiv e-prints, arXiv:1706.02808

\bibitem[{Owen \& Rudolf(2021)}]{owen:strong}
Owen, A.~B. \& Rudolf, D. 2021, SIAM Rev., 63(2), 360

\bibitem[{Peebles(1980)}]{peebles:lss}
Peebles, P. J.~E. 1980, The Large Scale Structure of the {U}niverse (Princeton,
  New Jersey: Princeton University Press)

\bibitem[{{Peebles} \& {Hauser}(1974)}]{peebles:statisticalIII}
{Peebles}, P.~J.~E. \& {Hauser}, M.~G. 1974, Ap.\ J.\ Suppl., 28, 19

\bibitem[{{Ragagnin} {et~al.}(2017){Ragagnin}, {Dolag}, {Biffi}, {Cadolle Bel},
  {Hammer}, {Krukau}, {Petkova}, \& {Steinborn}}]{ragagnin:web}
{Ragagnin}, A., {Dolag}, K., {Biffi}, V., {et~al.} 2017, Astronomy and
  Computing, 20, 52

\bibitem[{Ripley(1976)}]{ripley:second-order}
Ripley, B.~D. 1976, J.\ Appl.\ Prob., 13, 255

\bibitem[{Ripley(1988)}]{ripley:spatial}
Ripley, B.~D. 1988, Statistical Inference For Spatial Processes (Cambridge:
  Cambridge University Press)

\bibitem[{Rivolo(1986)}]{rivolo:two-point}
Rivolo, A.~R. 1986, ApJ, 301, 70

\bibitem[{{Ross} {et~al.}(2020){Ross}, {Bautista}, {Tojeiro}, {Alam}, {Bailey},
  {Burtin}, {Comparat}, {Dawson}, {de Mattia}, {du Mas des Bourboux},
  {Gil-Mar{\'\i}n}, {Hou}, {Kong}, {Lyke}, {Mohammad}, {Moustakas}, {Mueller},
  {Myers}, {Percival}, {Raichoor}, {Rezaie}, {Seo}, {Smith}, {Tinker},
  {Zarrouk}, {Zhao}, {Zhao}, {Bizyaev}, {Brinkmann}, {Brownstein}, {Rosell},
  {Chabanier}, {Choi}, {Chuang}, {Cruz-Gonzalez}, {de la Macorra}, {de la
  Torre}, {Escoffier}, {Fromenteau}, {Higley}, {Jullo}, {Kneib}, {McLane},
  {Mu{\~n}oz-Guti{\'e}rrez}, {Neveux}, {Newman}, {Nitschelm},
  {Palanque-Delabrouille}, {Paviot}, {Pullen}, {Rossi}, {Ruhlmann-Kleider},
  {Schneider}, {Maga{\~n}a}, {Vivek}, \& {Zhang}}]{ross:complete}
{Ross}, A.~J., {Bautista}, J., {Tojeiro}, R., {et~al.} 2020, Mon.\ Not.\ Roy.\
  Astron.\ Soc., 498, 2354

\bibitem[{{Ross} {et~al.}(2012){Ross}, {Percival}, {S{\'a}nchez}, {Samushia},
  {Ho}, {Kazin}, {Manera}, {Reid}, {White}, {Tojeiro}, {McBride}, {Xu}, {Wake},
  {Strauss}, {Montesano}, {Swanson}, {Bailey}, {Bolton}, {Dorta}, {Eisenstein},
  {Guo}, {Hamilton}, {Nichol}, {Padmanabhan}, {Prada}, {Schlegel},
  {Maga{\~n}a}, {Zehavi}, {Blanton}, {Bizyaev}, {Brewington}, {Cuesta},
  {Malanushenko}, {Malanushenko}, {Oravetz}, {Parejko}, {Pan}, {Schneider},
  {Shelden}, {Simmons}, {Snedden}, \& {Zhao}}]{ross:clustering}
{Ross}, A.~J., {Percival}, W.~J., {S{\'a}nchez}, A.~G., {et~al.} 2012, Mon.\
  Not.\ Roy.\ Astron.\ Soc., 424, 564

\bibitem[{{Saunders} {et~al.}(1992){Saunders}, {Rowan-Robinson}, \&
  {Lawrence}}]{saunders:spatial}
{Saunders}, W., {Rowan-Robinson}, M., \& {Lawrence}, A. 1992, \mnras, 258, 134

\bibitem[{{Shaw} {et~al.}(2021){Shaw}, {M{\o}ller}, \&
  {Waagepetersen}}]{shaw:globally}
{Shaw}, T., {M{\o}ller}, J., \& {Waagepetersen}, R. 2021, Australian \& New
  Zealand Journal of Statistics, 63, 93

\bibitem[{{Sinha} \& {Garrison}(2020)}]{sinha:corrfunc}
{Sinha}, M. \& {Garrison}, L.~H. 2020, MNRAS, 491, 3022

\bibitem[{Stoyan {et~al.}(1995)Stoyan, Kendall, \& Mecke}]{stoyan:stochgeom}
Stoyan, D., Kendall, W.~S., \& Mecke, J. 1995, Stochastic Geometry and its
  Applications, 2nd edn. (Chichester: John Wiley \& Sons)

\bibitem[{Stoyan \& Stoyan(1994)}]{stoyan:fractals}
Stoyan, D. \& Stoyan, H. 1994, Fractals, Random Shapes and Point Fields:
  Methods of Geometrical Statistics (Chichester: John Wiley \& Sons)

\bibitem[{Stoyan \& Stoyan(2000)}]{stoyan:improving}
Stoyan, D. \& Stoyan, H. 2000, Scandinavian Journal of Statistics, 27, 641

\bibitem[{Virtanen {et~al.}(2020)Virtanen, Gommers, Oliphant, Haberland, Reddy,
  Cournapeau, Burovski, Peterson, Weckesser, Bright, {van der Walt}, Brett,
  Wilson, Millman, Mayorov, Nelson, Jones, Kern, Larson, Carey, Polat, Feng,
  Moore, {VanderPlas}, Laxalde, Perktold, Cimrman, Henriksen, Quintero, Harris,
  Archibald, Ribeiro, Pedregosa, {van Mulbregt}, \& {SciPy 1.0
  Contributors}}]{virtanen:scipy}
Virtanen, P., Gommers, R., Oliphant, T.~E., {et~al.} 2020, Nature Methods, 17,
  261

\end{thebibliography}

\begin{appendix} 

\section{Expectation of pair-counts}
\label{sec:expectpair}

Estimators for the two-point density and the correlation function
using geometrical weights have been developed in spatial statistics
(e.g.\,\citealt{stoyan:fractals}).  These geometrical weights can be
derived with an application of the Campbell-Mecke formula to the
pair-counts.  These ideas can be found in several places, such as for
example \citet{ripley:second-order}, \citet{ohser:estimators},
\citet{fiksel:edge}, \citet{stoyan:stochgeom} and quite recently for
inhomogeneous point sets in \citet{shaw:globally}. We provide
analogous derivations to show the connection between the pair-counts
$\DR$ and $\RR$ and geometric quantities such as the set-covariance
and the area fraction (\citealt{kerscher:twopoint};
\citealt{stoyan:improving}).

The Campbell-Mecke formula connects the expectation value over
realisations of a point process $\Phi$ to integrals over $n$-point
densities (e.g.\,\citealt{stoyan:stochgeom}; \citealt{daley:introductionI}).
For suitable functions $f(\bfx)$ and $g(\bfx,\bfy),$ we have
\begin{align}
  \bbE\left[\sum_{\bfx\in\Phi} f(\bfx)\,\right]
  &= \int_{\bbR^3}f(\bfx)\, \varrho\, \rmd\bfx ,\\
  \bbE\left[ \sum_{\substack{\bfx,\bfy\in\Phi\\ \bfx\ne \bfy}} g(\bfx,\bfy)\,\right]
  &= \int_{\bbR^3}\int_{\bbR^3} g(\bfx,\bfy)\, \varrho_2(\bfx,\bfy)\, \rmd\bfx\rmd\bfy.
\end{align}
The Campbell-Mecke formula allows us to interchange the expectation of
sums with an integration over the number density $\varrho$ or the
two-point density $\varrho_2(\bfx,\bfy)$.
%
For a simple point process $\Phi,$ we define
\begin{align}
S &= \sum_{\substack{\bfx,\bfy\in\Phi\\ \bfx\ne \bfy}} 
\bbone_\CW(\bfx) \bbone_\CW(\bfy) \mDelta k_r^\mDelta(|\bfx-\bfy|) .
\end{align}
Considering the galaxy distribution as a realisation of a point
process $\Phi,$ we have $\DD=\frac{S}{N^2}$.
We apply the Campbell-Mecke formula to $\bbE[S]$, where we assume
homogeneity and isotropy
$\varrho_2(\bfx,\bfy)=\varrho_2(|\bfx-\bfy|)$.
\begin{align}
\label{eq:S-campbel-mecke}
\bbE[S] 
  &= \int_{\bbR^3}\int_{\bbR^3} \bbone_\CW(\bfx) \bbone_\CW(\bfy) \mDelta k_r^\mDelta(|\bfx-\bfy|)
  \varrho_2(|\bfx-\bfy|)\, \rmd\bfx\rmd\bfy  \nonumber \\
  &=  \int_{\bbR^3}
  \underbrace{\int_{\bbR^3} \bbone_\CW(\bfz+\bfy) \bbone_\CW(\bfy) \rmd\bfy}_{=\gamma_\CW(\bfz)} \
  \mDelta k_r^\mDelta(|\bfz|) \varrho_2(|\bfz|)\, \rmd\bfz  \nonumber \\
  &= \int_{0}^\infty 
  \underbrace{\int_0^\pi\!\!\! \int_0^{2\pi}\!\!\! \gamma_\CW(\bfz(s,\theta,\phi))
    \sin(\theta)\rmd\theta\rmd\phi}_{=4\pi\overline{\gamma_\CW}(s)} \
  \mDelta k_r^\mDelta(s) \varrho_2(s)\, s^2\rmd s \nonumber \\
  &= 4\pi \int_{r}^{r+\mDelta}\!\!  \overline{\gamma_\CW}(s)\ \varrho^2 (1+\xi(s))\ s^2\rmd s ,
\end{align}
with $\varrho_2(s)=\varrho^2(1+\xi(s))$ and the number density
$\varrho$ of the point process.

The random point set used in $\RR$ is a realisation of a Poisson
process with number density $\varrho_\text{r}=\frac{\Nr}{|\CW|}$ and
by definition a vanishing two-point correlation function
$\xi(r)=0$. Applying the Campbell-Mecke formula to
$\RR=\frac{S_\text{r}}{\Nr^2}$, we obtain the connection between the
pair-count $\RR$ and the isotropised set-covariance
(\citealt{kerscher:twopoint}; \citealt{stoyan:improving}):
\begin{align}
  \label{eq:RRt}
  \bbE[\RR(r)] = \RRt(r) & =  \frac{4\pi}{\Nr^2} \int_{r}^{r+\mDelta}\!\!
  \overline{\gamma_\CW}(s)\ \varrho_\text{r}^2 \ s^2\rmd s \\
  & \approx\ \frac{4\pi r^2 \mDelta}{|\CW|^2}\
  \overline{\gamma_\CW}(r) \quad \text{for } \mDelta \text{ small}.\nonumber 
\end{align}
Using the isotropised set covariance (see Appendix
\ref{sec:setcov-afrac-box}), we can calculate $\RRt(r)$ as a
geometrical reference value for rectangular boxes.

As mentioned above, we consider the galaxy distribution as a
realisation of a point process with number density
$\varrho=\frac{N}{|\CW|}$ and we seek to estimate its two-point
correlation function $\xi(r)$. From
\begin{align}
\frac{\bbE[\DD(r)]}{\bbE[\RR(r)]}
&= \frac{\Nr^2\ 4\pi\int_{r}^{r+\mDelta} \overline{\gamma_\CW}(s)\ \varrho^2(1+\xi(s)) s^2\rmd s }{
  N^2\ 4\pi\int_{r}^{r+\mDelta} \overline{\gamma_\CW}(s)\ \varrho_\text{r}^2 s^2\rmd s } \nonumber \\
&= 1 + \frac{\int_{r}^{r+\mDelta} \overline{\gamma_\CW}(s)\ \xi(s)\ s^2\rmd s }{
  \int_{r}^{r+\mDelta} \overline{\gamma_\CW}(s)\ s^2\rmd s }\\
& \approx\ 1 + \xi(r) \quad \text{for } \mDelta \text{ small},\nonumber 
\end{align}
we see that the estimator $\xi_\text{PH}(r)=
\frac{\DD(r)}{\RR(r)}-1$ of \cite{peebles:statisticalIII} is a `ratio
unbiased' estimator for the two-point correlation function $\xi(r)$.

To derive an analogous relation between $\DR$ and the average surface
area we consider
\begin{align}
T &= \sum_{i=1}^N \sum_{\bfy\in\Phi} 
\bbone_\CW(\bfy)\,  \mDelta k_r^\mDelta(|\bfx_i-\bfy|) ,
\end{align}
with a point process $\Phi$ and $\{\bfx_i\}_{i=1}^N$ being a given set of
points inside the sample geometry $\bfx_i\in\CW$.  With $\Phi,$ a
Poisson process, and using the Campbell-Mecke formula for
$\DR(r)=\frac{T}{N \Nr}$, we obtain
\begin{align}
  \bbE[T] &= \sum_{i=1}^N \int_{\bbR^3} \bbone_\CW(\bfy)\,
  \mDelta k_r^\mDelta(|\bfx_i-\bfy|)\, \varrho_\text{r}\, \rmd\bfy 
  = \frac{\Nr}{|\CW|} \sum_{i=1}^N  \Vt_r^\mDelta(\bfx_i),
\end{align}
where $\varrho_\text{r}=\frac{\Nr}{|\CW|}$ is the number density of
the Poisson process and
\begin{align}
  \label{eq:Vrdelta}
  \Vt_r^\mDelta(\bfx_i) 
  & = |S_r^\mDelta(\bfx_i)\cap\CW|
  = \int_r^{r+\mDelta}\!\!\!\!\! \text{area}(\partial\CB_{s}(\bfx_i)\cap\CW)\,\rmd s
\end{align}
is the volume of the spherical shell
$S_r^\mDelta(\bfx_i)=\{\bfy\in\bbR^3 \,|\, s < |\bfy-\bfx_i|\le
s+\mDelta \}$ with a radial range in $[r,r+\mDelta]$ around $\bfx_i$
inside the sample geometry $\CW$. For simple sample geometries, the
$\text{area}(\partial\CB_{s}(\bfx_i)\cap\CW)$ can be calculated
explicitly (see Appendix\,\ref{sec:setcov-afrac-box}). The integral in
Eq.\,(\ref{eq:Vrdelta}) can be evaluated using standard numerical
methods to obtain the expectation value of value $\DR(r)$:
\begin{align}
  \label{eq:DRt-area}
  \DRt(r)
  & = \frac{1}{|\CW|N}\sum_{i=1}^N \Vt_r^\mDelta(\bfx_i) \\
  & \approx\ \frac{1}{|\CW|}\, \frac{1}{N}
  \sum_{i=1}^N  \text{area}(\partial\CB_{r}(\bfx_i)\cap\CW)\,\mDelta
\quad \text{for } \mDelta \text{ small}. \nonumber
\end{align}
Clearly, the expectation value of the estimators from
\citet{davis:surveyV}, \citet{hewett:estimation}, and \citet{hamilton:towards}
can be expressed in terms of the istotropised set-covariance and the
average area fraction \citep{kerscher:twopoint}.

\subsection{Periodic boundaries}
\label{sec:periodic}

Let us assume that our window is a rectangular box
$\CW=[0,L_1]\times[0,L_2]\times[0,L_3]$ with periodic boundaries,
i.e. $\CW$ has the topology of a three-torus. The majority of
cosmological simulations enforce these boundary conditions. In such a
situation, no boundary corrections are needed in the calculation of the
two-point correlation function, but we have to respect the periodicity
in each coordinate direction \citep{stoyan:stochgeom}.  The distance
between two points $\bfx,\bfy\in\CW$ is
$d(\bfx,\bfy)=\sqrt{d_1^2+d_2^2+d_3^2}$, with
$d_i=\min\{|x_i-y_i|,L_i-|x_i-y_i|\}$.  This is applicable if $r$ is
smaller than any $L_i/2$.  For the shifted periodic box, we have
$\CW_{\bfx}=\{\bfy\, |\, \bfy-\bfx\in\CW\}=\CW$ and consequently the
(isotropised) set-covariance is constant
$\gamma_\CW(\bfx)=\overline{\gamma_\CW}(|\bfx|)=|\CW|$.
Similar to the derivation in Eq.\,(\ref{eq:S-campbel-mecke}), the
expectation value of $\DD(r)$ can be calculated using the
Campbell-Mecke formula:
\begin{align}
\bbE[\DD(r)] 
&= \frac{4\pi}{N^2}
\int_{r}^{r+\mDelta}\!\! |\CW|\, \varrho^2 (1+\xi(s))\, s^2\rmd s \nonumber \\
&= \frac{|S_r^\mDelta|}{|\CW|} +
\frac{4\pi}{|\CW|} \int_{r}^{r+\mDelta}\!\! \xi(s)\, s^2\rmd s \nonumber\\
& \approx\ \frac{4\pi r^2 \mDelta}{|\CW|} (1 + \xi(s))
\quad \text{for } \mDelta \text{ small}, 
\end{align}
where the volume of the shell
$|S_r^\mDelta|=\frac{4\pi}{3} \left((r+\mDelta)^3 - r^3\right)$.
Therefore, for small $\mDelta$, $\frac{|\CW|}{|S_r^\mDelta|}\DD(r)-1$ is
an unbiased estimate of the two-point correlation function $\xi(r)$ in
a periodic box; neither a random points set, nor a boundary
correction with geometric factors is needed.

\subsection{Expectation of pair-counts for inhomogeneous point sets.}
\label{sec:inhomogeneous}

As described in Sect.\,\ref{sec:incomplete} we are naturally
confronted with an inhomogeneous sampled galaxy distribution.  Here
we calculate the expectation values of $\DD', \DR'$ and $\RR'$ for
such an inhomogeneous situation.

One also has Campbell-Mecke formulas for an inhomogeneous point
process $\Phi'$ (\citealt{baddeley:nonandsemi}; \citealt{shaw:globally}):
\begin{align}
  \bbE\left[\sum_{\bfx\in\Phi'} f(\bfx)\,\right]
  &= \int_{\bbR^3}f(\bfx)\, \varrho'(\bfx)\, \rmd\bfx ,\\
  \bbE\left[ \sum_{\substack{\bfx,\bfy\in\Phi'\\ \bfx\ne \bfy}} g(\bfx,\bfy)\,\right]
  &= \int_{\bbR^3}\int_{\bbR^3} g(\bfx,\bfy)\, \varrho_2'(\bfx,\bfy)\, \rmd\bfx\rmd\bfy \\
  & \hspace{-0.5cm}= \int_{\bbR^3}\int_{\bbR^3} g(\bfx,\bfy)\,\varrho'(\bfx)\varrho'(\bfy)\,(1+\xi'(|\bfx-\bfy|))\, \rmd\bfx\rmd\bfy. \nonumber
\end{align}
Again we consider
\begin{align}
S' &= \sum_{\substack{\bfx,\bfy\in\Phi'\\ \bfx\ne \bfy}} 
\bbone_\CW(\bfx) \bbone_\CW(\bfy) \mDelta k_r^\mDelta(|\bfx-\bfy|) ,
\end{align}
and calculate the expectation (see \citealt{baddeley:nonandsemi} and
\citealt{shaw:globally} for similar derivations)
\begin{align}
\label{eq:S-campbel-mecke-prime}
\lefteqn{\bbE[S'] = \int_{\bbR^3}\!\int_{\bbR^3}\!\! \bbone_\CW(\bfx) \bbone_\CW(\bfy) \mDelta k_r^\mDelta(|\bfx-\bfy|)
  \varrho_2'(|\bfx-\bfy|)\, \rmd\bfx\rmd\bfy}  \nonumber \\
 &=  \int_{\bbR^3}\! \int_{\bbR^3}\!\! 
  \bbone_\CW(\bfz+\bfy) \bbone_\CW(\bfy) \varrho'(\bfz+\bfy)\varrho'(\bfy)  \rmd\bfy\
  \mDelta k_r^\mDelta(|\bfz|) (1+\xi'(|\bfz|)  \rmd\bfz  \nonumber \\
  %
  &= 4\pi \int_{r}^{r+\mDelta}\!\!  \overline{\Gamma}_\CW(s)\ (1+\xi'(s))\ s^2\rmd s ,
\end{align}
where $\overline{\Gamma}_\CW(s)$ is the density weighted isotropised
set-covariance
\begin{multline}
\label{eq:GammaW}
\overline{\Gamma}_\CW(s) = \frac{1}{4\pi}\int_0^\pi\!\!\! \int_0^{2\pi}\!\!\! \int_{\bbR^3}
  \bbone_\CW(\bfz(s,\theta,\phi)+\bfy) \bbone_\CW(\bfy) \\
  \varrho'(\bfz(s,\theta,\phi)+\bfy)\varrho'(\bfy)\ \rmd\bfy \sin(\theta)\rmd\theta\rmd\phi.
\end{multline}
For a homogeneous point distribution with $\varrho'(\bfx)=\varrho,$ we
consistently get
$\overline{\Gamma}_\CW(s)=\varrho^2\,\overline{\gamma_\CW}(s)$.

The random point set used in $\RR'$ is a realisation of an
inhomogeneous Poisson process with number density
$\tfrac{\Nr}{N}\varrho'(\bfx)$, and a vanishing two-point correlation
function $\xi'(r)=0$.  $N$ is the number of galaxies and $\Nr$ the
number of random points in $\CW$.  From
Eq.\,(\ref{eq:S-campbel-mecke-prime}) we get
\begin{align}
  \label{eq:RRtprime}
  \bbE[\RR'(r)] = \RRt'(r) & =  \frac{4\pi}{N^2} \int_{r}^{r+\mDelta} \overline{\Gamma_\CW}(s) \ s^2\rmd s \\
  & \approx\ \frac{4\pi r^2 \mDelta}{N^2}\
  \overline{\Gamma_\CW}(r) \quad \text{for } \mDelta \text{ small}.\nonumber 
\end{align}
Using Eq.\,(\ref{eq:S-campbel-mecke-prime}) and Eq.\,(\ref{eq:RRtprime}) we can determine 
\begin{align}
\frac{\bbE[\DD'(r)]}{\bbE[\RR'(r)]}
&= \frac{\int_{r}^{r+\mDelta} \overline{\Gamma_\CW}(s)\ (1+\xi'(s)) s^2\rmd s }{
  \int_{r}^{r+\mDelta} \overline{\Gamma_\CW}(s)\ s^2\rmd s } \nonumber \\
& \approx\ 1 + \xi'(r) \quad \text{for } \mDelta \text{ small},\nonumber 
\end{align}
Hence we arrive at the well-known result that the estimator
$\xi_\text{PH}'(r)= \frac{\DD'(r)}{\RR'(r)}-1$ of
\cite{peebles:statisticalIII} is `ratio unbiased'. $\xi_\text{PH}'(r)$
is an estimate of the two-point correlation function $\xi'(r)=\xi_G(r)$
if we apply the same selection effects to the random point set as we
find in the incompletely sampled galaxy distribution.

For a given point set $\{\bfx_i\}_{i=1}^N$ inside the sample geometry
$\bfx_i\in\CW,$ we calculate the expectation of $\DR'$.  Using the
Campbell-Mecke formula and with $\Phi'$ an inhomogeneous Poisson
process with number density $\tfrac{\Nr}{N}\varrho'(\bfx),$ we obtain
\begin{align}
  \label{eq:DRtprime}
  \bbE[\DR'] &= \DRt'(r) =
   \frac{1}{N\Nr} \sum_{i=1}^N \int_{\bbR^3} \bbone_\CW(\bfy)\,
  \mDelta k_r^\mDelta(|\bfx_i-\bfy|)\, \tfrac{\Nr}{N}\varrho'(\bfy)\, \rmd\bfy \nonumber\\
  & = \frac{1}{N^2} \sum_{i=1}^N  \Ut_r^\mDelta(\bfx_i),
\end{align}
where
\begin{align}
  \label{eq:Urdelta}
  \Ut_r^\mDelta(\bfx_i) 
  = \int_{\bbR^3} \bbone_\CW(\bfy)\,
  \mDelta k_r^\mDelta(|\bfx_i-\bfy|)\, \varrho'(\bfy)\, \rmd\bfy
\end{align}
is the integral of the density $\varrho'(\bfy)$ in the volume of the
spherical shell
$S_r^\mDelta(\bfx_i)=\{\bfy\in\bbR^3 \,|\, s < |\bfy-\bfx_i|\le
s+\mDelta \}$ inside $\CW$.

\section{Simple windows}
\label{sec:simple-win}

The area fraction $\text{area}(\partial\CB_r(\bfx)\cap \CW)$ and the
(isotropised) set covariance $\overline{\gamma_\CW}(r)$ can be
calculated for a rectangular box $\CW$.  We use these expressions to
determine the accuracy of the (quasi-)Monte~Carlo integration schemes.
We also reiterate the result for a sphere and for two-dimensions.

\subsection{Rectangular box}
\label{sec:setcov-afrac-box}

For a point $\bfx=(x_1,x_2,x_3)^T$ and a rectangular box
$\CW=[0,L_1]\times[0,L_2]\times[0,L_3]$ with side lengths $L_1>|x_1|,
L_2>|x_2|, L_3>|x_3|,$ the set--covariance is
\begin{equation}
  \label{eq:setcovbox}
  \gamma_\CW(\bfx) = (L_1-|x_1|)(L_2-|x_2|)(L_3-|x_3|),
\end{equation}
and  the isotropised set-covariance is for $r<\min\{L_1,L_2,L_3\}$
\begin{multline}
  \label{eq:isosetcovbox}
  \overline{\gamma_\CW}(r) = L_1 L_2 L_3 -\\
  -\frac{r}{2}(L_1 L_2+L_1 L_3+L_2 L_3) 
  + \frac{2r^2}{3\pi}(L_1+L_2+L_3) - \frac{r^3}{4\pi}; 
\end{multline}
(see e.g.\ \citealt{stoyan:fractals}; \citealt{kerscher:twopoint}).
A simple integration gives us the integrated isotropised
set-covariance:
\begin{multline}
  \label{eq:int_isosetcovbox}
  \int_r^R\overline{\gamma_\CW}(s) s^2\rmd s = 
  \tfrac{1}{3} L_1 L_2 L_3 (R^3 - r^3) -\\
  - \tfrac{1}{8}(L_1 L_2+L_1 L_3+L_2 L_3)(R^4 - r^4) +\\
  + \tfrac{2}{15\pi}(L_1+L_2+L_3) (R^5 - r^5)
  - \tfrac{1}{24\pi} (R^6 - r^6) ,
\end{multline}
which we need to calculate $\RRt(r)$ according to Eq.\,(\ref{eq:RRt}).

It is more involved to calculate the surface area of a sphere inside a
rectangular box $\CW$. Using the inclusion-exclusion formula,
{}\cite{baddeley:3dpoint} derive an explicit expression for the area
fraction $\text{area}(\partial\CB_r(\bfx)\cap \CW)$:
\begin{align}
  \lefteqn{\text{area}(\partial\CB_r(\bfx)\cap \CW)
  = 4\pi r^2 - \sum_{i=1}^3 \left\{A_1(x_i,r) + A_1(L_i-x_i,r)\right\} }\nonumber\\
  &+\sum_{i=1}^3 \sum_{j=i+1}^3 \Big\{\, A_2(x_i,x_j,r) + A_2(x_i,L_j-x_j,r) \nonumber\\[-1em]
  & \qquad \qquad\qquad
  + A_2(L_i-x_i,x_j,r) + A_2(L_i-x_i,L_j-x_j,r)\, \Big\} \nonumber\\
  &- A_3(x_1,x_2,x_3,r) - A_3(L_1-x_1,x_2,x_3,r)
  - A_3(x_1,L_2-x_2,x_3,r) \nonumber\\
  & - A_3(x_1,x_2,L_3-x_3,r) - A_3(L_1-x_1,L_2-x_2,x_3,r) \nonumber\\
  & - A_3(L_1-x_1,x_2,L_3-x_3,r) - A_3(x_1,L_2-x_2,L_3-x_3,r) \nonumber\\
  & - A_3(L_1-x_1,L_2-x_2,L_3-x_3,r) ,
\end{align}
with 
\begin{align}
A_1(t_1,r) &= 4 r^2 C(t_1/r,0,0)\nonumber\\
A_2(t_1,t_2,r) &= 2 r^2 C(t_1/r,t_2/r,0)\\
A_3(t_1,t_2,t_3,r) &= r^2 C(t_1/r,t_2/r,t_3/r).\nonumber
\end{align}
The following expression for $C(a,b,c)$ is almost the same as Eq.\,(37) from
{}\cite{baddeley:3dpoint}, but with two typos corrected.  These
correct results have already been used in the \texttt{spatstat}
package for \texttt{R} (\citealt{baddeley:spatstat}; see
\texttt{sphefrac.c} for the implementation).
For $a,b,c>0$ and $a^2+b^2+c^2<1,$ we get
\begin{align}
  \lefteqn{C(a,b,c) = \at\left(\frac{(1-a^2-c^2)^{1/2}}{ac}\right) +
  \at\left(\frac{(1-b^2-c^2)^{1/2}}{bc}\right)} \nonumber \\
  & + \at\left(\frac{(1-a^2-b^2)^{1/2}}{ab}\right) 
  - a \at\left(\frac{(1-a^2-c^2)^{1/2}}{c}\right)\nonumber \\
  & + a \at\left(\frac{b}{(1-a^2-b^2)^{1/2}}\right) 
  - b \at\left(\frac{(1-b^2-c^2)^{1/2}}{c}\right)\nonumber \\
  & + b \at\left(\frac{a}{(1-a^2-b^2)^{1/2}}\right)  
  - c \at\left(\frac{(1-a^2-c^2)^{1/2}}{a}\right)\nonumber \\
  & + c \at\left(\frac{b}{(1-b^2-c^2)^{1/2}}\right) - \pi .
\end{align}
For $a^2+b^2+c^2\ge1,$ we have $C(a,b,c)=0$. As special cases,
\begin{multline}
  C(a,b,0) = \at\left(\frac{(1-a^2-b^2)^{1/2}}{ab}\right)  
  -a \at\left(\frac{(1-a^2-b^2)^{1/2}}{b}\right) \\
  -b \at\left(\frac{(1-a^2-b^2)^{1/2}}{a}\right) ,
\end{multline} 
and $C(a,0,0)=\tfrac{\pi}{2}(1-a)$.

\subsection{Sphere}

For a spherical sample $\CW=\CB_R$ centred at the origin with radius
$R$ and with $r=|\bfx|<R,$ the set-covariance reads
\begin{equation}
\gamma_\CW(\bfx) = \frac{4\pi}{3}
\left(R^3 -\frac{3}{4}rR^2 + \frac{1}{16} r^3\right).
\end{equation}
Due to symmetry, the isotropised set-covariance  is simply
$\overline{\gamma_\CW}(|\bfx|)=\gamma_\CW(\bfx)$. 
The area fraction for a point $\bfq\in\CW$ at a distance
$|\bfq|=s>R-r$ from the origin, and with $r<R$ is
$\text{area}(\partial\CB_r(\bfq)\cap \CW) = 4\,\alpha\, r^2$, with
$\cos(\alpha) = \frac{r^2+s^2-R^2}{2rs}$.

\subsection{Two dimensions}

Here we give the expressions for the set-covariance in
two dimensions. For a point $\bfx=(x_1,x_2)^T\in\mathbb{R}^2$ and a
rectangle $\CW=[0,L_1]\times[0,L_2]$ with side lengths
$L_1>|x_1|, L_2>|x_2|,$ the set--covariance is
\begin{equation}
   \gamma_\CW(\bfx)= (L_1-|x_1|)(L_2-|x_2|),
\end{equation}
and for $r<\min\{L_1,L_2\},$ the isotropised set-covariance is
\citep{ripley:spatial}
\begin{align}
  \overline{\gamma_\CW}(r)
  & = \frac{1}{2\pi}\int_0^{2\pi}\gamma_\CW(\bfx(r,\phi))\,\rmd\phi \nonumber \\
  & = L_1 L_2 -\frac{2r}{\pi}(L_1+L_2) + \frac{r^2}{\pi}.
\end{align}
The analogue to $\text{area}(\partial\CB_r(\bfx)\cap \CW)$ is the
arc-length $\text{arc}(C_r(\bfx)\cap W)$ of the circumference of a
circle $C_r(\bfx)$ with radius $r$ centred on a point $\bfx\in W$
inside the sample window $W$.  It can be calculated from the
intersection points of the circle with the sides of $W$ and the angles
between them and the centre \citep{stoyan:stochgeom}.

\section{Implementation and run time}
\label{sec:implementation}

We use Python with NumPy \citep{harris:numpy} and SciPy
\citep{virtanen:scipy} to implement the data processing, calculate
geometric quantities, generate random points, and to generate
randomised Halton sequences (provided in SciPy starting with version
$\ge$\texttt{1.7.0}).
The performance critical parts are the pair-counting and a function
for checking whether or not points are inside the observational
window.  We implemented these routines as C++ functions parallelised
using OpenMP \citep{dagum:openmp}. They are made callable from Python
via pybind11 \citep{jakob:pybind11}.
In the computations of $\DD$ and $\RR,$ we obviously avoid the double
counting by using
$\sum_i\sum_{j\ne i}\ldots = 2\sum_i\sum_{j>i}\ldots$. We implement
the double sums without using a tree, resulting in a code with a run
time quadratic in the number of points (see below). Using OpenMP, we
could achieve a reduction of the computing time almost linear with the
number of computing cores used (at least up to the 20 cores we were
able to access; see also \citealt{alonso:cute}).
Only these optimisations are included in our code but perhaps it
is versatile enough to help with future implementations\footnote{You find a
basic version of the code at
\url{https://homepages.physik.uni-muenchen.de/\~Martin.Kerscher/software/accuratexi/}.}.

\subsection{Run time}

As initial guidance, the run time of the pair-count calculations
should scale with the number of distance calculations actually
performed. This again depends on the number of (quasi-)random or data
points, and on the data structure and algorithm used.
With our direct implementation of the double sums from
eqs.\,(\ref{eq:defDD}), (\ref{eq:defDR}), (\ref{eq:defRR}),
(\ref{eq:RR2x3d}) we expect the run time to scale as
$T_{\DD}\propto N^2$, $T_{\DR}\propto N\Nr$, $T_{\RR}\propto \Nr^2$
and $T_{\RR,{2x3d}}\propto \Nq^2$, respectively.  This is independent
of the number of radius intervals, because we can sort the counts into
the corresponding radius bin.
The calculation of $\DR_{\text{shell}}$ should scale as
$T_{\DR,{\text{shell}}}\propto N\Nsh$, but we need to
repeat the pair-count calculation for each radius bin. One could
envision optimising this for convex windows.

\begin{figure}
\begin{center}
\includegraphics[width=0.38\textwidth]{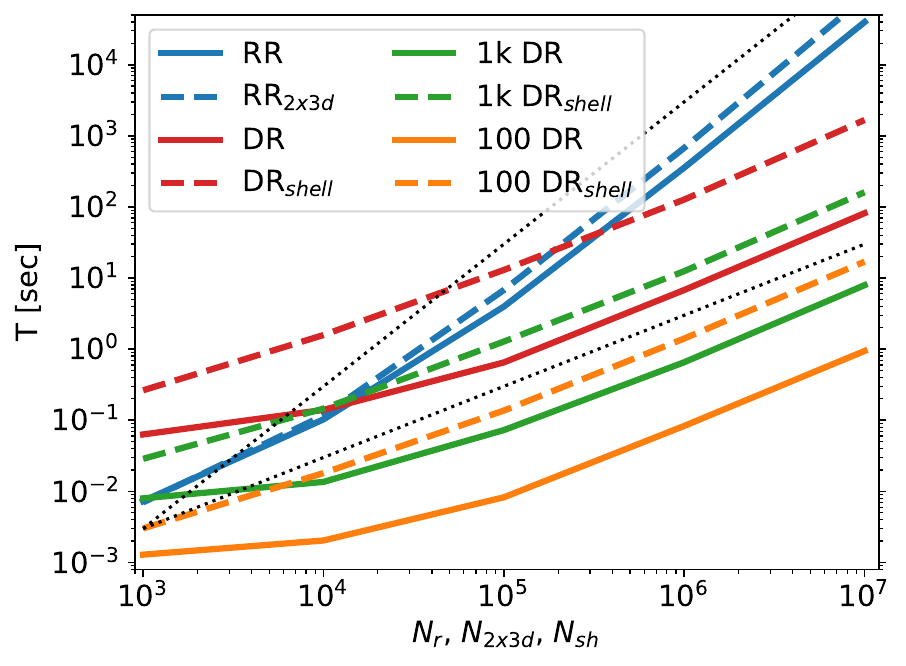}
\end{center}
\caption{
  \label{fig:run time_paircounts}
  Run time of the pair-counts for $r\in[5,10]\hMpc$ and  $W=[0,1]^3$.
  The black dotted lines are proportional to $\Nr$ and
  $\Nr^2$. }
\end{figure}

To estimate the run time, we ran our mildly optimised code on a current
small workstation equipped with an Intel Xeon W1350 processor with six
cores and a clock frequency of at least 3.3\,GHz.  Memory was not an
issue in these calculations.
As expected, we see from Fig.\,\ref{fig:run time_paircounts} that for
large $N_r$, the run time of $\RR$ and $\RR_{\text{2x3d}}$ grows
quadratically with $N_r$, and that $\RR_{\text{2x3d}}$ is slower than
$\RR$ by a factor of two.
The run time of $\DR$ and $\DR_{\text{shell}}$ shows the expected
linear scaling with $\Nr$. At fixed $\Nr$ and $N$, the run time of
$\DR_{\text{shell}}$ is more than one order of magnitude longer than
$\DR$. But we note that for fixed accuracy we need significantly less
quasi-random points in $\DR_{\text{shell}}$.
We also consider subsamples with $N=1000$ and $N=100$ data points from
the $N=10429$ clusters and confirm the linear scaling of $T_{\DR}$ and
$T_{\DR,{\text{shell}}}$ with $N$.  We find the same results for
$r\in[30,35]\,\hMpc$ and $r\in[100,105]\,\hMpc$.

\begin{figure}
\begin{center}
\includegraphics[width=0.38\textwidth]{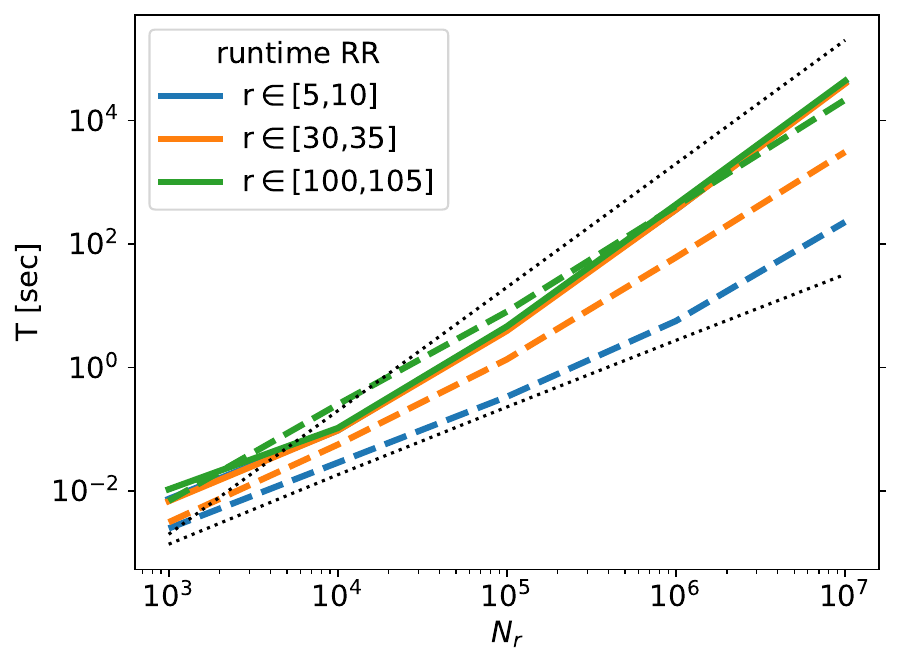}
\end{center}
\caption{
  \label{fig:run time_RRtree}
  Run time of $\RR$ in $[0,1]^3$ for the direct double sum
  implementation (three solid lines, almost on top of each other), and
  the implementation using \texttt{KDTree} (dashed lines).  The black
  dotted lines are proportional to $\Nr\log(\Nr)$ and $\Nr^2$.}
\end{figure}

In all our calculations, we used a parallelised direct double sum
implementation to calculate the pair-counts.  As already mentioned, a
significant speedup can be expected if one uses tree-based methods
\citep{moore:fast}.
To illustrate this, we compare the run time of our parallelised
pair-count implementation for $\RR$ to an implementation using the
kd--tree provided with SciPy (\texttt{KDTree}).
As the direct double sum implementation ran parallelised and the
\texttt{KDTree} on a single core, the absolute run times are not
comparable, but the scaling with $\Nr$ is the essential result. 
The direct implementation shows the expected scaling $\propto \Nr^2$
of the run time in Fig.\,\ref{fig:run time_RRtree}.  For the
tree-based method, we observe that for small radii the run time scales
with $\propto \Nr\log(\Nr)$ and a significant speedup can be achieved.
However, for large radii, the run time scales $\propto \Nr^2$,
comparable to the direct implementation.
In a similar comparison for the window $[0,1]\times[0,0.1]^2$ the
run time of the tree based method scales $\propto \Nr^2$ for all radii.
We therefore confirm a significant speedup for tree-based methods,
especially in voluminous samples and for small radii.
Tree-based methods, or the other computational optimisations
mentioned in the introduction, can be applied directly to $\DD$,
$\RR$, $\DR$, \emph{and} $\RR_{\text{2x3d}}$, and should speedup these
calculations further.
Whether or not similar optimisation schemes can be found for
$\DR_{\text{shell}}$ is an open question.

For a given point set and window there is plenty of scope for
optimising the run times. However, such an optimisation problem easily
turns into a multi-parameter endeavour.  The run time not only depends
the accuracy goal, the radius, the bin width, and the density of the point
set but also on the hardware platform, the operating system, the
algorithms, the data structures, the details of the implementation,
and so on.

\end{appendix}

\end{document}